\documentclass[aps,prd,notitlepage,showpacs,nofootinbib,superscriptaddress]{revtex4-1}

\usepackage{graphicx}

\usepackage[utf8]{inputenc} 

\usepackage{amsmath}
\usepackage{amssymb}
\usepackage{float}
\usepackage{comment}
\usepackage{slashed}
\usepackage[normalem]{ulem}
\usepackage{booktabs}
\usepackage{array}
\usepackage{multirow}
\usepackage{bm}
\usepackage{mathrsfs}

\usepackage{caption}
\usepackage{subcaption}
\usepackage{soul}

\usepackage{amsmath}
\usepackage{amssymb}

\usepackage[usenames,dvipsnames]{color}
\usepackage[colorinlistoftodos]{todonotes}
\usepackage[colorlinks=true,citecolor=darkred,urlcolor=darkred, pdfborder={0 0 0}]{hyperref}
\usepackage[normalem]{ulem}

\definecolor{darkred}{rgb}{0.6,0,0}
\usepackage[colorinlistoftodos]{todonotes}

\definecolor{linkcolor}{rgb}{0,0,0.5}

\newcommand\bpar[1]{\overset{%
   \scalebox{0.5}{$\bm{(\mkern-1mu-\mkern-1mu)}$}}{#1}}

 \def\three{\ensuremath{\mathbf{3}}}
 \def\threeS{\ensuremath{\mathbf{3^*}}}	

\def\gsim{\raise0.3ex\hbox{$\;>$\kern-0.75em\raise-1.1ex\hbox{$\sim\;$}}}
\def\lsim{\raise0.3ex\hbox{$\;<$\kern-0.75em\raise-1.1ex\hbox{$\sim\;$}}}

\newcommand{\sm}{{Standard Model }}

\usepackage{soul}

\definecolor{mightnightblue}{RGB}{25,25,112}

\definecolor{brown}{rgb}{0.59, 0.29, 0.0}

\newcommand {\ignore}[1]{}

\def\vev#1{\left\langle #1\right\rangle}

\def\21{$\mathrm{SU(2)_L \otimes U(1)_Y}$}

\def\sm{Standard Model }

\providecommand{\be}{ \begin{equation} } 
\providecommand{\ee}{ \end{equation} }
\providecommand{\bea}{\begin{eqnarray}}
\providecommand{\eea}{\end{eqnarray}}

\providecommand{\to}{\rightarrow}

\newcommand{\AddrIFIC}{%
  Institut de F\'{i}sica Corpuscular --
  C.S.I.C./Universitat de Val\`{e}ncia \\ Parc Cient\'ific de Paterna.
 C/ Catedr\'atico Jos\'e Beltr\'an, 2 \\ E-46980 Paterna (Valencia) - SPAIN}
\begin{document}

\title{\boldmath \color{BrickRed} Flavor-changing axions
and Dirac neutrino masses}

\author{Anirban Karan}\email{anirban1karan@gmail.com}
\affiliation{\AddrIFIC}
\author{Julio Leite}\email{juliorafa@gmail.com }
\affiliation{\AddrIFIC}

\author{Jos\'{e} W. F. Valle}\email{valle@ific.uv.es}
\affiliation{\AddrIFIC}

\begin{abstract}
\vskip 1cm 
Implementing the axion concept in the context of 3-3-1 extensions of the Standard Model (SM) leads to richer properties than in the simpler axion setups, and related to the Dirac neutrino seesaw mechanism. In this way the smallness of neutrino masses, the strong CP problem, the nature of dark matter and the number of families all have a common origin.
Besides having an enhanced coupling to photons, our revamped axion can also be distinguished from DFSZ and KSVZ axions through its couplings to fermions.
The latter lead to 
interesting phenomenological consequences, including flavor-changing axion-emitting two-body K, B and D meson decays. 
\end{abstract}
\maketitle
\noindent

\section{Introduction}\label{sec:intro}

Originally proposed as a solution to the strong CP problem in quantum chromodynamics (QCD)~\cite{Weinberg:1977ma,Wilczek:1977pj,Peccei:1977hh}
the axion also leads to an attractive candidate for the cold dark matter in the universe~\cite{Preskill:1982cy,Abbott:1982af,Dine:1982ah}.
In its simplest realizations, axion models leave aside the issue of neutrino mass generation (a recent exception is given in Ref.~{\cite{Berbig:2022pye}}), now required in view of the historical discovery of neutrino oscillations~\cite{Kajita:2016cak,McDonald:2016ixn} that require two nonvanishing neutrino mass splittings.

We recall that the question of why one has three families of fundamental fermions in nature is nicely explained within the $\mathrm{SU(3)_c}\otimes\mathrm{SU(3)_L}\otimes\mathrm{U(1)_X}$ framework~\cite{Singer:1980sw,Valle:1983dk,Pisano:1991ee,Frampton:1992wt}. In this setup, this follows from the peculiar way in which gauge anomalies cancel in 3-3-1 models.
Addressing the strong CP problem by implementing the axion solution requires an extended 3-3-1 scalar sector~\cite{Pal:1994ba,Dias:2003iq}.
Here we adopt the framework provided by 3-3-1 extensions of the standard model, where the neutrino mass generation, the strong CP problem and the number of families are all addressed simultaneously.

More specifically, we consider versions that contain two extra $\mathrm{U(1)}$ factors, one associated with the B-L symmetry, that leads to the Dirac nature of neutrinos, and the other related to the anomalous Peccei-Quinn (PQ) symmetry that leads to the axion. 
{We explore the phenomenology of the axion within the 3-3-1 framework proposed in Ref.~\cite{Dias:2020kbj}.}
The large PQ symmetry breaking scale ensures not only the invisibility of the axion~\cite{Dine:1981rt,Zhitnitsky:1980tq,Kim:1979if,Shifman:1979if}, which can play the role of cold dark matter, but also explains the smallness of neutrino masses. 
However, despite the large value of the Peccei-Quinn symmetry breaking scale, our scenario can lead to experimentally detectable axion phenomena.

Besides an enhanced coupling to photons, our axion field combines features of DFSZ and KSVZ axions, as it couples both to SM and to the extra exotic quarks expected in the 3-3-1 scenario. 
 In the relevant limit the axion behaves as a (flavored) DFSZ-like axion. 
 Thus, $\mathrm{SU(3)_L}$ effectively leads to non-trivial flavor-changing neutral currents (FCNCs) associated to new interactions, such as those involving the axion. The study of how these flavor-violating quark interactions can induce new axion-emitting K, B and D decay modes is the main goal of this paper.

\section{Model Overview}\label{sec:model} 

As our basic setup we consider the SVS model~\cite{Singer:1980sw}, an electroweak extension of the SM built upon the 3-3-1 gauge group, {\it i.e.} $\mathrm{SU(3)_c}\otimes\mathrm{SU(3)_L}\otimes\mathrm{U(1)_X}$.
In order to account for neutrino masses and solve the strong CP problem in an inter-connected manner, we further extend the group structure by adding a $\mathrm{U(1)_N}\otimes \mathrm{U(1)_{PQ}}$ symmetry.  The first Abelian group is free from anomalies and closely related to the B-L symmetry, while the second group is an anomalous Peccei-Quinn symmetry that will be associated with the axion field. 

The particle content and the corresponding symmetry properties of fermions and scalars  are summarized in Table \ref{tab1}. 
The left-handed fermions are assigned to the fundamental and anti-fundamental representations of $\mathrm{SU(3)_L}$ and can be decomposed as 
\bea\label{eq:lhfer}
\psi_{aL}=\left( \nu_{aL}, e_{aL},(\nu_{aR})^{c}\right)^{T}\,, \qquad\qquad
Q_{\alpha L}=\left( d_{\alpha L},-u_{\alpha L},D_{\alpha L}\right)^{T}\,,\qquad\quad
Q_{3L}=\left( u_{3L}, d_{3L},U_{3L}\right)^{T}\,,
\eea
with $a=1,2,3$ and $\alpha=1,2$. Except for the left-handed neutrinos, $\nu_{aL}$, whose right-handed partners, $\nu_{aR}$, are in the same $\mathrm{SU(3)_L}$ triplet~\cite{Valle:1983dk}, all left-handed fermions in Eq. (\ref{eq:lhfer}) have right-handed $\mathrm{SU(3)_L}$-singlet partners, see Table \ref{tab1}. 
\begin{table}[ht!] 
	\begin{centering}
     {\renewcommand{\arraystretch}{1.4}
		\begin{tabular}{|c|c|c|}
            \hline
			\toprule 
			\,\,Field\,\, &\,\,\,\,\,\, 3-3-1-1\,\,\,\,\,\, &\,\,$\mathrm{U(1)_{PQ}}$\,\,\,
			\tabularnewline
			\hline\hline
			\midrule
			$\psi_{aL}$ & $\left(\mathbf{1},\three,-\frac{1}{3},-\frac{1}{3}\right)$ &\,\,\,\,$\frac{1}{2}(-x_{\sigma}+x_{1}+x_{3})$\,\,  \tabularnewline
			$e_{aR}$ & $\left(\mathbf{1},\mathbf{1},-1,-1\right)$ &$\frac{1}{2}(x_{\sigma}+3x_{1}+3x_{3})$ 
   \tabularnewline
			$Q_{\alpha L}$ & $\left(\mathbf{3},\threeS,0,-\frac{1}{3}\right)$ & $x_{Q_{\alpha}}$
   \tabularnewline
			$Q_{3L}$ & $\left(\mathbf{3},\three,\frac{1}{3},1\right)$ & $x_{Q_{\alpha}}-x_{\sigma}-x_{3}$
     \tabularnewline
			$u_{a R}$ & $\left(\mathbf{3},\mathbf{1},\frac{2}{3},\frac{1}{3}\right)$ &\,\,\,$x_{Q_{\alpha}}-(x_{\sigma}+x_{1}+x_{3})$
   \tabularnewline
			$ U_{3R}$ & $\left(\mathbf{3},\mathbf{1},\frac{2}{3},\frac{7}{3}\right)$ &$x_{Q_{\alpha}}-x_{\sigma}-2x_{3}$ 
   \tabularnewline
			$d_{a R}$ & $\left(\mathbf{3},\mathbf{1},-\frac{1}{3},\frac{1}{3}\right)$ &$x_{Q_{\alpha}}+x_{1}$
      \tabularnewline
            $D_{\alpha R}$ & $\left(\mathbf{3},\mathbf{1},-\frac{1}{3},-\frac{5}{3}\right)$ & $x_{Q_{\alpha}}+x_{3}$
            \tabularnewline
            \hline
            $S_{aL}$ & $\left(\mathbf{1},\mathbf{1},0,-1\right)$  &$\frac{1}{2}(x_{\sigma}-x_{1}+x_{3})$
           \tabularnewline
            $S_{aR}$ & $\left(\mathbf{1},\mathbf{1},0,-1\right)$   & $\frac{1}{2}(-x_{\sigma}-x_{1}+x_{3})$
            \tabularnewline
            \hline \hline
			$\Phi_{1}$ & $\left(\mathbf{1},\three,-\frac{1}{3},\frac{2}{3}\right)$ &$x_{1}$
   \tabularnewline
			$\Phi_2$ & $\left(\mathbf{1},\three,\frac{2}{3},\frac{2}{3}\right)$ & $-(x_{\sigma}+x_{1}+x_{3})$
   \tabularnewline
			$\Phi_3$ & $\left(\mathbf{1},\three,-\frac{1}{3},-\frac{4}{3}\right)$ &$x_{3}$
   \tabularnewline
			\hline
			\,\,\,$\sigma$\,\,\, & \,\,\,$\left(\mathbf{1},\mathbf{1},0,0\right)$\,\,\, & $x_{\sigma}$
 \tabularnewline
			\bottomrule
			\hline
		\end{tabular}}
		\par
             \end{centering}
		\caption{
        Field content and symmetry transformation rules. The second column, labeled 3-3-1-1, shows the  transformation properties under the anomaly-free symmetry group $\mathrm{SU(3)_C}\otimes\mathrm{SU(3)_L}\otimes\mathrm{U(1)_X}\otimes\mathrm{U(1)_N}$.}\label{tab1}
\end{table}
 
Note that we introduce pairs of neutral fermions, $S_{aL,R}$, vector-like under all symmetries except for $\mathrm{U(1)_{PQ}}$. 
In the scalar sector the model includes three $\mathrm{SU(3)_L}$ triplets and a singlet, 
$\Phi_1=(\phi_1^0,\, \phi_1^-,\, \widetilde\phi_1^0)^T$, $\Phi_2=(\phi_2^+,\, \phi_2^0,\, \widetilde\phi_2^+)^T$, $\Phi_3=(\phi_3^0,\, \phi_3^-, \,\widetilde\phi_3^0)^T$ and $\sigma$, which, after spontaneous symmetry breaking, can be written as 
\begin{equation}
\label{eq:scalars}
\Phi_1=\left(
\begin{array}{c}
\frac{v_1+s_1+i a_1}{\sqrt{2}}\\
\phi^{-}_{1}\\
\widetilde{\phi}^{0}_{1}
\end{array}
\right),\qquad
\Phi_2=\left(
\begin{array}{c}
\phi_2^{+}\\
\frac{v_2+s_2+i a_2}{\sqrt{2}}\\
\widetilde{\phi}_2^+
\end{array}
\right),\qquad
\Phi_3=\left(
\begin{array}{c}
 \phi_3^{0} \\
 \phi_3^- \\
 \frac{w+s_3+i a_3}{\sqrt{2}} \\
\end{array}
\right),\qquad
\sigma= \frac{v_\sigma+s_\sigma+i a_\sigma}{\sqrt{2}}\,,
\end{equation}
where $v_\sigma$ characterizes the PQ breaking scale,
$w$ the $\mathrm{SU(3)_L}$ breaking scale, while $v_1,~v_2$ denote the electroweak scale, with $v_\sigma \gg w \gg \sqrt{v_1^2+v_2^2} \equiv v_{EW}$.~\footnote{It is worth mentioning that the components $\widetilde \phi_i^0$ should not be confused with the charge conjugate of scalar fields.} \\[-.4cm] 

The symmetry-breaking pattern (omitting the conserved  $\mathrm{SU(3)_c}$ factor for short), can be summarized as:
{\small
\be\label{eq:ssbpattern}
\mathrm{SU(3)_L}\otimes\mathrm{U(1)_X}\otimes\mathrm{U(1)_N}\otimes \mathrm{U(1)_{PQ}} \xrightarrow{v_\sigma} \mathrm{SU(3)_L}\otimes\mathrm{U(1)_X}\otimes\mathrm{U(1)_N} 
\xrightarrow{w}\mathrm{SU(2)_L}\otimes\mathrm{U(1)_Y}\otimes\mathrm{U(1)_{B-L}}
\xrightarrow{v_1,v_2} \mathrm{U(1)_Q}\otimes\mathrm{U(1)_{B-L}}\,.
\ee 
}
The residual subgroups, associated with the electric $Q$ and $B-L$ charges, are generated by 
\bea\label{eq:QBL}
    Q = T_3 - \frac{1}{\sqrt{3}} T_8 + X~\quad\quad \mbox{and}\quad\quad
    B-L = -\frac{4}{\sqrt{3}} T_8 + N~,
\eea
where $T_i$ are the $\mathrm{SU(3)_L}$ generators, whereas $X$ and $N$ generate $\mathrm{U(1)_X}$ and $\mathrm{U(1)_N}$, respectively. The $\mathrm{U(1)_N}$ can either be chosen as a global or gauge symmetry. For simplicity, we assume that the anomaly-free $\mathrm{U(1)_{N}}$ is a global symmetry. We also use the defining/fundamental representation of $\mathrm{SU(3)_L}$ generators with $T_i$ equal to half of the corresponding Gell-Mann matrix. Thus the diagonal generators $T_3$ and $T_8$ become $T_3=\text{diag}\left(\frac{1}{2},-\frac{1}{2},0\right)$ and $T_8=\text{diag}\left(\frac{1}{2\sqrt 3},\frac{1}{2\sqrt 3},-\frac{1}{\sqrt 3}\right)$. In this representation, the first and third components of any triplet will posses the same electric charge, while the second component will have the charge lesser by one unit. Moreover, the first and second components of triplets (and anti-triplets) will share the same $B-L$ charge.

We stress that introducing the neutral fermions $S_{aL,R}$ and the scalar $\sigma$, absent in the original SVS model, brings deep consequences. As we discuss next, this extended scenario allows for an intriguing connection between seemingly unrelated issues: the smallness of neutrino masses, their Dirac nature, the strong CP problem and the nature of cold dark matter.

\subsection{Vector Boson Spectrum}\label{sec:GaugeSpectrum} 

To derive the vector boson spectrum, we define how the covariant derivative acts on the scalar triplets $\Phi_i$:
\begin{equation}
	D_\mu \Phi_i= \left[ \partial_\mu - i g_L T^a W^{a}_{\mu}- i g_X X B_{\mu}\right]\Phi_i = \left(\partial_\mu- i \frac{g_L}{2}\mathcal{P}_\mu\right)\Phi_i\,,
\end{equation}
where $W^{a}_\mu$ denote the $\mathrm{SU(3)_L}$ gauge fields, while $B_\mu$ is the $\mathrm{U(1)_X}$ gauge field. Here $\mathcal{P}_\mu$ is written in a matrix form as 
\begin{eqnarray}
\mathcal{P}_\mu=
	\begin{pmatrix}
	W^3_\mu+\frac{W^8_\mu}{\sqrt{3}} + 2 t_X X B_\mu & \sqrt{2}W_\mu^+ & \sqrt{2}X_\mu^{0} \\ \sqrt{2} W_\mu^- & -W_\mu^3+\frac{W_\mu^8}{\sqrt{3}} + 2t_X X B_\mu & \sqrt{2}W_\mu^{\prime-} \\
	\sqrt{2} X_\mu^{0*} &\sqrt{2} W_\mu^{\prime+} &  2\left(-\frac{W_\mu^8}{\sqrt{3}} + t_X X B_\mu\right) \end{pmatrix},
\end{eqnarray}
with $t_X = g_X/g_L$ and
\begin{equation}\label{cgb}
 W_\mu^\pm = \frac{W_\mu^1\mp i W_\mu^2}{\sqrt{2}}\,,\quad\quad X_\mu^{0(*)}=\frac{W_\mu^4\mp i W_\mu^5}{\sqrt{2}}\quad\quad\mbox{and}\quad\quad  W_\mu^{\prime \pm}=\frac{W_\mu^6\pm i W_\mu^7}{\sqrt{2}}\,.
\end{equation} 

After spontaneous symmetry breaking all the gauge bosons, except for the photon, become massive. 
The charged fields, $W_\mu^\pm$ and $W_\mu^{\prime\pm}$, remain unmixed due to their different $B-L$ charges.
Their masses are given by
\begin{equation}\label{eq:Wmasses}
m^2_W=\frac{g_L^2}{4}  \left(v_1^2+v_2^2\right) = \left(\frac{g_L v_{EW}}{2}\right)^2 \quad\quad\mbox{and}\quad\quad
m^2_{W^{\prime}}=\frac{g_L^2}{4} \left(v_2^2+w^2\right),
\end{equation}
where $W_\mu$ is identified with the \sm $W$ boson.

The complex gauge boson $X^0_\mu$, with $B-L=2$, does not mix with the other neutral vector fields, $W^3_\mu, W^8_\mu, B_\mu$ since for such fields $B-L = 0$. The $X^0_\mu$ mass is then 
\begin{equation}
m^2_{X^0}=\frac{g_L^2}{4}  \left(v_1^2+w^2\right).
\end{equation}

Finally, the real neutral gauge bosons mix after  spontaneous symmetry breaking (SSB), and their mass matrix, in the basis $V_\mu^0 = (W^3_\mu, W^8_\mu, B_\mu)$, can be written as 
\begin{equation}
M^2_{V}=\frac{g_L^2}{2}\left(
\begin{array}{cccc}
 \frac{1}{2} \left(v_1^2+v_2^2\right)  & \frac{v_1^2-v_2^2 }{2 \sqrt{3}} & -\frac{1}{3} \left(v_1^2+2v_2^2\right)  t_X \\
 \frac{v_1^2-v_2^2 }{2 \sqrt{3}} & \frac{1}{6} \left(v_1^2+v_2^2+4 w^2\right) &  \frac{\left(-v_1^2+2 v_2^2+2 w^2\right)}{3 \sqrt{3}}\,t_X \\
 -\frac{1}{3} \left( v_1^2+2v_2^2\right)  t_X & \frac{\left(-v_1^2+2 v_2^2+2 w^2\right)  }{3 \sqrt{3}}\,t_X & \frac{2}{9} \left( v_1^2+4v_2^2+w^2\right)  t_X^2  \\
\end{array}
\right)\,.
\end{equation}
The diagonalization of $M^2_V$ can be performed by unitary transformations, $U_V M_V^2 U_V^T$, defined as 
\bea\label{eq:UN}
U_V =U_V^2\,U_V^1
&=& \left(
\begin{array}{ccc}
 1 & 0 & 0 \\
 0 & c_Z & s_Z \\
 0 & -s_Z & c_Z \\
\end{array}
\right)\,\left(
\begin{array}{ccc}
 \frac{\sqrt{3} t_X}{\sqrt{4 t_X^2+3}} & -\frac{t_X}{\sqrt{4 t_X^2+3}} & \frac{\sqrt{3}}{\sqrt{4 t_X^2+3}} \\
 \frac{\sqrt{t_X^2+3}}{\sqrt{4 t_X^2+3}} & \frac{\sqrt{3} t_X^2}{\sqrt{\left(t_X^2+3\right) \left(4 t_X^2+3\right)}} & -\frac{3 t_X}{\sqrt{\left(t_X^2+3\right) \left(4 t_X^2+3\right)}} \\
 0 & \frac{\sqrt{3}}{\sqrt{t_X^2+3}} & \frac{t_X}{\sqrt{t_X^2+3}} \\
\end{array}
\right)\,,
\eea
\bea\label{eq:thetaZ}
\text{with } \quad
\tan 2 \theta_Z 
&=&\frac{3 \sqrt{4 t_X^2+3} \left[\left(2 t_X^2-3\right) v_1^2+\left(4 t_X^2+3\right) v_2^2\right]}{\left(2 t_X^4-24 t_X^2-9\right) v_1^2+\left(8 t_X^4-6 t_X^2-9\right) v_2^2 +2 \left(t_X^2+3\right)^2 w^2}\,.
\eea
 
The first transformation, $U_V^1$, singles out the massless state, the photon, while the second one diagonalizes the resulting $2\times 2$ block matrix parametrized by the mixing angle $\theta_Z$ between the two massive states $Z^1$ and $Z^2$. The resulting masses are
\bea
m_{Z^{1,2}}^2 &=& \frac{g_L^2}{18}  \left[\left(t_X^2+3\right) (v_1^2+w^2)+\left(4 t_X^2+3\right) v_2^2\right.\nonumber\\
&&\left.\mp \sqrt{[\left(t_X^2+3\right) (v_1^2+w^2)+\left(4 t_X^2+3\right) v_2^2]^2-9 \left(4 t_X^2+3\right) \left(v_1^2 v_2^2+v_1^2w^2+v_2^2 w^2\right)}\right]\,.
\eea
In the limit $w\gg v_1^2+v_2^2\equiv v_{EW}$, we have 
\bea \label{eq:Zmasses}
m_{Z^1}^2 &\simeq& \frac{g_L^2 \left(3+4t_X^2 \right)(v_1^2 + v_2^2)}{4 \left(3+t_X^2 \right)}\quad\quad \mbox{and}\quad\quad
m_{Z^2}^2 \simeq \frac{g^2_L\left(t_X^2+3\right)}{9}   w^2\,, \\
\mbox{with}\quad\theta_{Z} &\simeq& \frac{3 \sqrt{4 t_X^2+3} \left[\left(2 t_X^2-3\right) v_1^2+\left(4 t_X^2+3\right) v_2^2\right]}{4 \left(t_X^2+3\right)^2 w^2}\ll 1\,.\nonumber
\eea 
Using the relation $s_W = \sqrt{3}t_X/\sqrt{3+4t_X^2}$ between the parameter $t_X = g_X/g_L$ and the Weinberg angle, $\theta_W$, as derived in the next section, we find that the SM prediction $m_{Z^1} \simeq m_W/c_W$ for $Z^1_\mu \simeq Z^{SM}_\mu $ is recovered in this limit.

\subsection{Scalar Spectrum} 
\label{sec:scalar}

As already mentioned, this model contains three complex scalar triplets and one complex singlet. 
After spontaneous symmetry breaking, these generate the following neutral scalar, pseudoscalar and charged scalar components:
\begin{itemize}
    \item Four neutral scalars $\{s_1,\, s_2,\, s_3,\, s_\sigma\}$ mixing together to form four scalar mass-eigenstates $\{H_1,\, H_2,\, H_3,\, H_4\}$. 
    \item Four pseudoscalars $\{a_1,\, a_2,\, a_3,\, a_\sigma\}$ mixing together to produce the light axion $a$, a heavy pseudoscalar $A$ and two Goldstone bosons $\{G_1,\, G_2\}$.  
    \item The remaining two neutral complex scalars, i.e. $\widetilde\phi_1^0$ and $\phi_3^0$ (its conjugate), form one heavy complex scalar $\varphi^0$ and a complex Goldstone boson $G_3$.
    \item In the charged scalar sector, $\phi_2^\pm$ mixes with $\phi_1^\pm$, leading to a heavy scalar $H_1^\pm$ along with the Goldstone boson $G_4^\pm$. Similarly, $\widetilde\phi_2^\pm$ mixes with $\phi_3^\pm$ to produce a heavy scalar $H_2^\pm$ along with the Goldstone boson $G_5^\pm$.
\end{itemize}
The detailed mass spectrum of the scalars follows from the scalar potential based on the content of Table \ref{tab1}, and can be expressed as 
\begin{eqnarray}\label{V}
V &=& \,
 \sum_{i=1}^{3}\left[\mu_i^2 \Phi^{\dagger}_i\Phi_i+\lambda_{i}(\Phi^{\dagger}_i\Phi_i)^2\right]+\sum_{i<j}^3\left[\lambda_{ij}(\Phi^{\dagger}_i\Phi_i)(\Phi^{\dagger}_j\Phi_j)+\tilde{\lambda}_{ij}(\Phi^{\dagger}_i\Phi_j)(\Phi^{\dagger}_j\Phi_i)
\right]\,\\
 &+& \,
  \sum_{i}^3 \lambda_{i\sigma}(\Phi^{\dagger}_i\Phi_i)(\sigma^*\sigma)+
 \mu_\sigma^2 \sigma^*\sigma+\lambda_\sigma(\sigma^*\sigma)^2 -\left(\lambda_A\, \sigma\,\Phi_1\Phi_2\Phi_3+\mathrm{h.c.}\right)\,.\nonumber
\end{eqnarray}

From the scalar potential above and the field decomposition in Eq. (\ref{eq:scalars}), we extract the following tadpole equations 
\begin{eqnarray}
 v_1 \left( 2 \mu_1^2 + 2 \lambda_{1} v_1^2 + \lambda_{12} v_2^2  + \lambda_{13} w^2   + \lambda_{1\sigma} v_\sigma^2 \right) &=& \lambda_A v_2 w  v_\sigma\,,\\
 v_2 \left( 2 \mu_2^2 +  2 \lambda_2 v_2^2  + \lambda_{12} v_1^2  + \lambda_{23} w^2   + \lambda_{2\sigma} v_\sigma^2   \right) &=& \lambda_A v_1 w v_\sigma \,,\nonumber\\
 w \left( 2 \mu_3^2 + \lambda_{13} v_1^2  + \lambda_{23} v_2^2 + 2 \lambda_{3} w^2 + \lambda_{3\sigma} v_\sigma^2  \right) &=& \lambda_A v_1 v_2 v_\sigma\,,\nonumber\\
 v_\sigma \left( 2 \mu_\sigma^2 + \lambda_{1\sigma} v_1^2 + \lambda_{2\sigma} v_2^2 + \lambda_{3\sigma} w^2 + 2 \lambda_{\sigma}  v_\sigma^2 \right) &=& \lambda_A v_1 v_2 w\,,\nonumber
\end{eqnarray}
which we solve simultaneously for the dimensionful constants $\mu_1,\mu_2, \mu_3$ and $\mu_\sigma$. 

We now calculate the tree-level scalar spectrum, starting from the CP-odd scalars. When grouped in the basis $(a_1, a_2, a_3, a_\sigma )$ these  share the squared mass matrix 
\begin{eqnarray}
M_a^2 = \frac{\lambda_A}{2}\left(
\begin{array}{cccc}
 \frac{v_2 w v_\sigma}{v_1} &  w v_\sigma & v_2 v_\sigma & v_2 w \\
  w v_\sigma & \frac{v_1 w v_\sigma}{v_2} & v_1 v_\sigma & v_1 w \\
 v_2 v_\sigma & v_1 v_\sigma & \frac{v_1 v_2 v_\sigma}{w} & v_1 v_2 \\
 v_2 w & v_1 w & v_1 v_2 & \frac{v_1 v_2 w}{v_\sigma} \\
\end{array}
\right)\,.
\end{eqnarray}
After diagonalization, we find only one massive CP-odd state at tree-level 
\begin{equation}\label{A}
 A = \frac{1}{\sqrt{N_A}}\left[  v_2 w v_\sigma\, a_1 + v_1 w v_\sigma\, a_2 + v_1 v_2 v_\sigma \,a_3 +  v_1v_2w\,a_\sigma \right]\,,
\end{equation}
\begin{equation}
 \text{where } \quad N_A=v_1^2 v_2^2 w^2+ v_\sigma^2( v_1^2 v_2^2  + v^2_{EW} w^2)\,,
\end{equation}
and whose mass is given by  
\begin{equation}
 m_A^2 = \lambda_A \frac{ v_1^2 v_2^2 w^2+ v_\sigma^2( v_1^2 v_2^2  + v^2_{EW} w^2)}{2 v_1 v_2 w v_\sigma}  \,.
\label{eq:mA}
\end{equation}  
Two other CP-odd eigenstates, $G_1$ and $G_2$, are would-be Goldstone bosons absorbed by the neutral vector bosons, $Z^1_\mu$ and $Z^2_\mu$, of the extended gauge sector. 
Finally, the remaining CP-odd state is the axion field associated with the spontaneous breaking of the anomalous $\mathrm{U(1)_{PQ}}$ symmetry~\footnote{
This explicit axion profile determination can also be obtained just from symmetry using the Noether's theorem method given in~Ref.~\cite{Schechter:1981cv}.} 
\begin{equation}\label{eq:axion}
 a = \frac{1}{\sqrt{N_a}}\left[ -v_1 v_2^2 w^2\, a_1 - v_2 v_1^2 w^2\, a_2 - w v_1^2 v_2^2\, a_3 + v_\sigma(v_1^2 v_2^2+v_{EW}^2 w^2 )\,a_\sigma \right]\,,
\end{equation}
where the normalization constant $N_a$ is given by 
\begin{equation}\label{Na}
 N_a=\left(v_1^2 v_2^2 + v_{EW}^2 w^2\right)N_A\,.
\end{equation}
One sees that, in the limit of interest, $v_\sigma\gg w \gg v_1,v_2$, the axion is mainly the imaginary part of 
$\sigma$.\\[-.2cm]  

Turning now to the CP-even scalars, in the basis $(s_1, s_2, s_3, s_\sigma )$, the relevant squared mass matrix is given by 
\begin{eqnarray}
M_s^2 = \frac{1}{2}\left(
\begin{array}{cccc}
 4\, \lambda_{1} v_1^2+\frac{\lambda_A v_2 w v_\sigma }{ v_1} & 2\lambda_{12} v_1 v_2 -\lambda_A w v_\sigma & 2\lambda_{13} v_1 w  - \lambda_A v_2 v_\sigma & 2\lambda_{1 \sigma} v_1 v_\sigma -\lambda_A v_2 w \\
 2\lambda_{12} v_1 v_2-\lambda_A w v_\sigma & 4\, \lambda_{2} v_2^2 +\frac{\lambda_A v_1 v_\sigma w}{ v_2} &  2v_2 w \lambda_{23}-\lambda_A v_1 v_\sigma  &  2\lambda_{2 \sigma} v_2 v_\sigma -\lambda_A v_1 w\\
 2 \lambda_{13} v_1 w -\lambda_A v_2 v_\sigma & 2\lambda_{23}v_2 w -\lambda_A v_1 v_\sigma & 4\,  \lambda_{3} w^2 + \frac{\lambda_A v_1 v_2 v_\sigma}{ w} & 2\lambda_{3 \sigma}w v_\sigma -\lambda_A v_1 v_2 \\
 2\lambda_{1 \sigma} v_1 v_\sigma-\lambda_A v_2 w & 2\lambda_{2 \sigma} v_2 v_\sigma -\lambda_A v_1 w & 2\lambda_{3 \sigma} w v_\sigma  - \lambda_A v_1 v_2 &4 \lambda_\sigma  v_\sigma^2 + \frac{\lambda_A v_1 v_2 w}{ v_\sigma} \\
\end{array}
\right)\,.
\label{eq:Ms}
\end{eqnarray}

In general, the matrix above leads to four non-vanishing eigenvalues, associated to four massive scalar bosons, $H_1$, $H_2$, $H_3$ and $H_4$.
For $v_\sigma= 10^{12}$ GeV, $w= 10^{4}$ GeV, and $\sqrt{v_1^2+v_2^2}= 246$ GeV, the heavier state is $H_4 \simeq s_\sigma$, which becomes much heavier than the others, $m_{H_4}^2 \simeq 2 \lambda_\sigma v_\sigma^2 $, and hence decouples from the rest. 
One of the lighter states is identified with the $125$ GeV Higgs boson, $H_1 \equiv h$. 

In addition to these neutral scalars, the model contains the complex neutral fields $\widetilde{\phi}_1^0$ and $\phi_3^{0}$ which have opposite $B-L$ charge (see Table \ref{tab1}), and when grouped in the basis $(\widetilde{\phi}_1^0,\phi_3^{0*} )$, share the following squared mass matrix 
\begin{eqnarray}
M_{\phi^0}^2 = \frac{1}{2}\left(
\begin{array}{cc}
 w \left(\tilde{\lambda}_{13} w +\frac{\lambda_A v_2 v_\sigma}{v_1}\right) & \lambda_Av_2 v_\sigma + \tilde{\lambda}_{13} v_1 w  \\ \lambda_A v_2 v_\sigma + \tilde{\lambda}_{13} v_1 w  & v_1 \left(\tilde{\lambda}_{13} v_1 +\frac{\lambda_A v_2 v_\sigma}{w}\right) \\
\end{array}
\right)\,.
\end{eqnarray}
In the mass basis, only one of the  states appears in the physical spectrum
\begin{eqnarray}
\varphi^0=\frac{w \widetilde{\phi}_1^0+v_1 \phi_3^{0*}}{\sqrt{v_1^2+w^2}},
\end{eqnarray}
and is heavy, with squared mass
\begin{equation}
m_{\varphi^0}^2=\frac{(v_1^2+w^2) (\tilde{\lambda}_{13} v_1 w+\lambda_A v_2 v_\sigma)}{2 v_1 w}\,.
\label{eq:mphi0}
\end{equation}
The other state $G_3$, orthogonal to $\varphi^0$, is massless and absorbed by the gauge sector. 

Finally, writing the charged scalars in the basis $(\phi_2^\pm,\phi_1^\pm,\widetilde{\phi}_2^\pm,\phi_3^\pm )$, we find the squared mass matrix
\begin{eqnarray}
M_\pm^2 = \frac{1}{2}\left(
\begin{array}{cccc}
 v_1 \left(\tilde{\lambda}_{12}v_1 +\frac{\lambda_A w v_\sigma }{v_2}\right) & \lambda_A w v_\sigma + \tilde{\lambda}_{12} v_1 v_2  & 0 & 0 \\
 \lambda_A w v_\sigma + \tilde{\lambda}_{12} v_1 v_2 & v_2 \left(\tilde{\lambda}_{12} v_2 +\frac{\lambda_A w v_\sigma }{v_1v_2w}\right) & 0 & 0 \\
 0 & 0 & w \left( \tilde{\lambda}_{23} w +\frac{\lambda_A v_1 v_\sigma}{v_2}\right) & \lambda_A v_1 v_\sigma + \tilde{\lambda}_{23} v_2 w  \\
 0 & 0 & \lambda_A v_1 v_\sigma + \tilde{\lambda}_{23} v_2 w & v_2 \left( \tilde{\lambda}_{23}v_2 +\frac{\lambda_A v_1 v_\sigma}{w}\right) \\
\end{array}
\right)\,.
\end{eqnarray}

As expected from Table \ref{tab1}, the charged fields with different $B-L$ charges do not mix.
Diagonalizing the matrix above, we find two massive charged scalar fields
\begin{equation}
H^{\pm}_1= \frac{v_1 \phi^{\pm}_2+v_2 \phi^{\pm}_1}{\sqrt{v_1^2+v_2^2}}\,,\quad\quad H^{\pm}_2=\frac{w\widetilde{\phi}_2^{\pm}+v_2 \phi^{\pm}_3  }{\sqrt{v_2^2 + w^2}}\,,
\end{equation}
whose masses are
\begin{eqnarray}
m_{H_1^\pm}^2&=&\frac{\left(v_1^2+v_2^2\right) (\tilde{\lambda}_{12}v_1 v_2 +\lambda_A w v_\sigma)}{2 v_1 v_2}\,\quad\quad\mbox{and}\quad\quad
m_{H_2^\pm}^2 = \frac{\left(v_2^2+w^2\right) (\tilde{\lambda}_{23}v_2 w +\lambda_A v_1 v_\sigma)}{2 v_2 w}\, ,
\label{eq:charged}
\end{eqnarray}
while the other two  massless states, $G_4^\pm$ and $G_5^\pm$, are absorbed by the charged gauge boson sector. 

\subsection{Fermion Masses }\label{sec:lept-mass-mixings}

The Yukawa Lagrangian for the leptons can be written as
\begin{eqnarray}\label{lagYl}
 -\mathcal{L}_{\rm Yl} &=& \,
 y^{e}_{ab}\,\overline{\psi_{aL} }  \,\Phi_2 e_{bR}  + y^{\nu_1}_{ab} \,\overline{ \psi_{aL}}\,\Phi_1 \,S_{bR} + y^{\nu_2}_{ab} \,\overline{ \psi_{aL}}\,\Phi_3 \, (S_{bL})^c +
 y^{S}_{ab}\, \overline{S_{aL}} \,\sigma\, S_{bR}  + \mathrm{h.c.}\,.
\end{eqnarray}
leading to the following mass matrices.

\vspace{.3cm}
\noindent {\it \underline{Charged Lepton masses} }\\ 

The charged lepton masses emerge from the first term when $\Phi_2$ acquires a vev
\begin{equation}\label{eq:clmm}
 M_{e} = \frac{y^e v_2}{\sqrt{2}}\,,
\end{equation}
where the family indices have been omitted. Notice that in the above equation, the index $e$ is used to represent collectively all charged leptons, not to be mistaken for the electron only.

\vspace{.3cm}
\noindent {\it \underline{Seesaw-suppressed Dirac neutrino masses} }\\ 

\begin{figure}[htbp] 
\begin{center}
\includegraphics[scale=1.2,height=3.5cm]{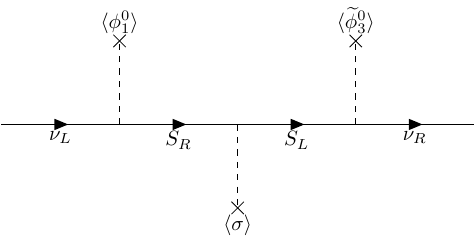}
\caption{
Type-I Dirac seesaw mechanism suppressed by the Peccei-Quinn scale  $\vev{\sigma}\equiv v_\sigma$.  }
\label{DiracSeesawDia}\label{fig}
\end{center}
\end{figure}

On the other hand, neutrino masses are generated via the type-I Dirac seesaw mechanism { \footnote{For 3-3-1 neutrino mass generation schemes employing the type-II Dirac seesaw mechanism see, for example, Refs. \cite{Angel:2025luo,Reig:2016ewy}.}}, illustrated in Fig. \ref{DiracSeesawDia}. 

Choosing $N=(\nu, S)$ as the basis, the neutral mass matrix is given as 
\begin{eqnarray}\label{ssm}
 M_{N} =\frac{1}{\sqrt{2}}\begin{pmatrix} 
       0 & y^{\nu_1} v_1 \\ (y^{\nu_2})^T w & y^{S} v_\sigma
      \end{pmatrix}\,.
\end{eqnarray}
where the heavy mass term $\overline{N_L} M_{N} N_R$ arises from the PQ-scale $v_\sigma$. 
In the limit $v_\sigma\gg w \gg v_{EW}$, we can make use of the (Dirac) seesaw expansion~\cite{Schechter:1981cv} to obtain the light neutrino mass matrix 
\begin{equation} \label{ssm2}
m_\nu^D \simeq \frac{ y^{\nu_1} (y^S)^{-1} (y^{\nu_2})^T }{\sqrt{2}}   \frac{v_1 w}{v_\sigma}\,.
\end{equation}
One sees that neutrino masses are suppressed by the ratio $w/v_\sigma$ with respect to the EW scale. 
Therefore, in order to  
account for the small neutrino masses, a relatively low 3-3-1 scale is required, compared to the Peccei-Quinn scale $v_\sigma \sim 10^{12}$ GeV. Note that this is consistent with a viable axion cold dark matter picture.
This implies that the rich 3-3-1 phenomenology could be within foreseeable experimental reach. \\

Turning now to the quark Yukawa Lagrangian, it can be expressed as
\begin{eqnarray}\label{lagYq}
 -\mathcal{L}_{\rm Yq} &=& \,
y^{u}_{\alpha a}\, \overline{Q_L^{\alpha}}\,\Phi_2^{*} \, u^{a}_R   +  
y^{u}_{3a}\, \overline{Q_L^{3}}\, \Phi_1 \,u^{a}_R  
+ y^{d}_{3a} \, \overline{Q_L^{3}}\, \Phi_2 \, d^{a}_R  + 
y^{d}_{\alpha a} \, \overline{Q_L^{\alpha}}\,\Phi_1^{*} \,d^{a}_R  \nonumber \\
&+& 
 y^{U}_{33}\, \overline{Q_L^{3}} \,\Phi_3  \,U^3_R
  +  y^{D}_{\alpha\beta} \, \overline{Q_L^{\alpha}}\,\Phi_3^{*}\,D_R^{\beta}  
 + \mathrm{h.c.}
\end{eqnarray}

\vspace{.3cm}
\noindent {\it \underline{Quark masses} }\\ 

Following spontaneous symmetry breaking, the up-type quark mass matrix in the basis $(u_{a},U_3)$ is given as
\begin{equation}\label{uqmass}
M_{u}=\frac{1}{\sqrt{2}}\left(
\begin{array}{cccc}
 -v_2 y^u_{11} & -v_2  y^u_{12} &  -v_2  y^u_{13} & 0 \\
 -v_2 y^u_{21} &  -v_2 y^u_{22} &  -v_2  y^u_{23} & 0 \\
 v_1 y^u_{31} &  v_1  y^u_{32} &  v_1  y^u_{33} & 0 \\
0 & 0 & 0 & w y^{U}_{33} \\
\end{array}
\right)=\left(
\begin{array}{cccc}
 m^u_{3\times 3} &  0_{3\times 1} \\
0_{1\times 3} & \frac{w y^{U}_{33}}{\sqrt{2}} \\
\end{array}
\right)\,.
\end{equation}

Similarly, for the down-type quarks, in the basis $(d_{a},D_{\alpha})$, we have 
\begin{equation}\label{dqmass}
\begin{split}
M_{d}&=\frac{1}{\sqrt{2}}\left(
\begin{array}{ccccc}
 v_1 y^d_{11} & v_1  y^d_{12} &  v_1  y^d_{13} & 0 &0\\
  v_1 y^d_{21} &   v_1 y^d_{22} &  v_1  y^d_{23} & 0 &0\\
  v_2  y^d_{31} &   v_2  y^d_{32} &  v_2  y^d_{33} & 0 &0\\
0 & 0 & 0 &  w y_{11}^{D} & w y_{12}^{D} \\
0 & 0 & 0 &  w y_{12}^{D} &  w y_{22}^{D} \\
\end{array}
\right)=\left(
\begin{array}{cccc}
 m^d_{3\times 3} &  0_{3\times 2} \\
0_{2\times 3} & m^D_{2\times 2} \\
\end{array}
\right).
\end{split}
\end{equation}
The block-diagonal structure of the above matrices follows from the conservation of $U(1)_{B-L}$ which forbids the mixing between light and heavy quarks.

\section{Fermion gauge interactions}
\label{sec:Fermion-mixing}

We now briefly discuss the structure of the charged current weak interactions of quarks and leptons.

\subsection{Charged current interactions}\label{subsec:CCs}

In the weak basis, the fermions interact with the electrically charged gauge bosons via
\be
\mathcal{L}_{CC} = \frac{g_L}{\sqrt{2}} W^+_\mu\left[ \overline{\nu_{aL}}\gamma^\mu e_{aL} + \overline{u_{aL}} \gamma^\mu d_{aL} \right] + \frac{g_L}{\sqrt{2}} W^{\prime+}_\mu \left[ \overline{\left(\nu_{aR}\right)^c} \gamma^\mu e_{aL} + \overline{u_{\alpha L}} \gamma^\mu \mathcal{D}_{\alpha L} + \overline{\mathcal{U}_{3 L}} \gamma^\mu d_{3 L} \right] + \mbox{h.c.}
\ee 
where the first term is identical to the SM interaction\footnote{ Similarly, the interactions involving the electrically neutral complex gauge boson, $X^0_\mu$, can be written as $\mathcal{L}_{X} = (g_L/\sqrt{2}) X_\mu^0\, [ \overline{\nu_{aL}} \gamma^\mu \left(\nu_{aR}\right)^c - \overline{\mathcal{D}_{\alpha L} } \gamma^\mu d_{\alpha L}  + \overline{u_{3 L}} \gamma^\mu \mathcal{U}_{3 L}] + \mbox{h.c.}$.}.

\vspace{.2cm}
\noindent {\it \underline{Lepton mixing matrix} }\\[-2mm] 

Implementing the transformations on the lepton fields required to diagonalise the charged lepton and light neutrino mass matrices in Eqs. (\ref{eq:clmm}) and (\ref{ssm2}), namely $U_{L}^e$ and $ U_{L}^\nu$, respectively, leads to the form the lepton mixing matrix 
\begin{equation}
  \label{eq:lepton-mixing}
   V_{LEP}=(U_{L}^e)^\dagger U_{L}^\nu\,.
\end{equation}
The charged lepton piece is unitary, while the neutrino part comes as a truncated rectangular matrix~\cite{Schechter:1980gr} due to the mixing with the heavy mediators $S_a$ with masses proportional to the PQ scale $v_\sigma$. 
As a result the effective mixing matrix relevant for describing neutrino propagation is only nearly unitary.

\vspace{.2cm}
\noindent {\it \underline{Quark CKM mixing matrix} } \\[-2mm] 

The standard $3\times 3$ up-type (down-type) mass matrices are diagonalized by rotating the left- and right-handed quark fields to the mass basis via the unitary transformations $U_{L,R}^{u}$ ($U_{L,R}^{d}$).  
Since the $B-L$ symmetry forbids mixing between standard and exotic quarks, the CKM matrix describing light quark mixing remains unitary and defined, as usual, as 
\begin{equation}
  \label{eq:CKM}
  V_{CKM}=(U_{L}^u)^\dagger U_{L}^d\,.
\end{equation}

\subsection{Neutral current interactions}\label{subsec:FCNCs}

The interactions between a neutral gauge boson, $\widetilde V^0_\mu\,\,(= A_\mu, Z^1_\mu, Z^2_\mu)$, and two fermions, $f_i$, can be expressed as 
\bea\label{eq:Bf}
\mathcal{L}_{\widetilde V,\,f} &=& 
\widetilde V^0_\mu \sum_{i,j}
\overline{f_{i}}\gamma^\mu\left[\left(g_{\widetilde V,\,f}^{V}\right)_{ij}-\left(g_{\widetilde V,\,f}^{A}\right)_{ij}\gamma^5\right] f_{j}\,.
\eea 
For the case of charged leptons ($f=e$) the present model leads to, 
\be
 \left(g_{\widetilde V,\,e}^{V(A)}\right)_{ij}  = \frac{g_{ \widetilde V,\,e_{L} } \pm g_{\widetilde V,\,e_{R} }}{2}\delta_{ij}\nonumber\,,
\ee 
in the mass basis. The coefficients are shown in Table \ref{tab:coeff}, and we have defined the electric charge, $e$, in terms of the Weinberg angle, $\theta_W = \theta_W(t_X)$, as
\be\label{eq:e}
e = g_L s_W = g_L \frac{\sqrt{3} t_X}{\sqrt{3+4t_X^2}}\,,
\ee
with $s_W = \sin{\theta_W}$. Notice that, in this case, all interactions are flavor-conserving. \\[-.2cm] 

In contrast, for the case of quarks, the interactionswith neutral gauge bosons become more involved, since one generation of left-handed quarks ($Q_{3L}$) comes in a different group representation from the other two ($Q_{\alpha L}$).  
These couplings have both flavor-conserving as well as flavor-violating contributions.
The coefficients for $q=u,d$ (mass basis) in Eq. (\ref{eq:Bf}) take the form 
\be
\left(g_{\widetilde V,\,q}^{V(A)}\right)_{ij}  = \frac{ \left(g_{\widetilde V,\,q_{\alpha L} } \pm g_{\widetilde V,\,q_{a R} }\right)\delta_{ij} + g_{\widetilde V,\,q}^{FV} (X^q)_{ij}}{2},
\ee
where the FV superscript denotes the flavor-violating piece, and
\be\label{eq:Xq}
g_{\widetilde V,\,q}^{FV} =g_{\widetilde V,\,q_{\alpha L}}-g_{\widetilde V,\,q_{3L}} \quad \mbox{and} \quad  X^{q} =\left[(U^q_L)^\dagger\mbox{diag}\left(0,0,-1\right)U^q_L\right]\,,
\ee 
and they are given in Table \ref{tab:coeff}. 
\begin{table}[h!]
    \centering
    \renewcommand{\arraystretch}{1.8} 

    \begin{minipage}{0.55\textwidth} 
        \centering
        \begin{tabular}{|c||c|c|c|}
            \hline
           \,\,\,$g_{B,f}$\,\,\,  & $Z^1_\mu$ & $Z^2_\mu$ & \,\,\,$A_\mu$\,\,\,\\ \hline\hline
            $e_{aL}$ & $\frac{g_L}{c_{W}} \left(-\frac{1}{2} + s_{W}^2\right) C_1$ & $\frac{g_L}{c_{W}} \left(-\frac{1}{2} + s_{W}^2\right) C_2$ & $-e$ \\ 
            \cline{1-3}
            $e_{aR}$ &  $\frac{g_L}{c_{W}} s_{W}^2 C_1$ & $\frac{g_L}{c_{W}} s_{W}^2 C_2$ & \\ \hline\hline
            $d_{\alpha L}$ &  $\frac{g_L}{c_{W}} \left(-\frac{1}{2} + \frac{1}{3}s_{W}\right) C_3$ & $\frac{g_L}{c_{W}} \left(-\frac{1}{2} + \frac{1}{3}s_{W}\right) C_4$ & \\
            \cline{1-3}
            $d_{3 L}$ & $\frac{g_L}{c_{W}} \left(-\frac{1}{2} + \frac{1}{3}s_{W}\right) C_1$ & $\frac{g_L}{c_{W}} \left(-\frac{1}{2} + \frac{1}{3}s_{W}\right) C_2$ & $-\frac{1}{3} e$ \\  
            \cline{1-3}
            $d_{aR}$ &  $\frac{g_L}{c_{W}} \left(\frac{1}{3}s_{W}\right) C_1$ & $\frac{g_L}{c_{W}} \left(\frac{1}{3}s_{W}\right) C_2$ &\\ \hline\hline
            $u_{\alpha L}$ &  $\frac{g_L}{c_{W}} \left(\frac{1}{2} - \frac{2}{3}s_{W}\right) C_1$ & $\frac{g_L}{c_{W}} \left(\frac{1}{2} - \frac{2}{3}s_{W}\right) C_2$ & \\ 
            \cline{1-3}
            $u_{3 L}$  & $\frac{g_L}{c_{W}} \left(\frac{1}{2} - \frac{2}{3}s_{W}\right) C_5$ & $\frac{g_L}{c_{W}} \left(\frac{1}{2} - \frac{2}{3}s_{W}\right) C_6$ & $\frac{2}{3} e$ \\ 
            \cline{1-3}
            $u_{aR}$ &  $\frac{g_L}{c_{W}} \left(-\frac{2}{3}s_{W}\right) C_1$ & $\frac{g_L}{c_{W}} \left(-\frac{2}{3}s_{W}\right) C_2$ & \\ \hline
        \end{tabular}
    \end{minipage}%
    \hspace{-1cm} 
    \begin{minipage}{0.4\textwidth} 
        \centering
        \begin{tabular}{|c|c|}
            \hline
            $C_1$ & $c_{Z} - \frac{1}{\sqrt{3 - 4 s_{W}^2}} s_{Z}$ \\ \hline
            $C_2$ & $-\frac{1}{\sqrt{3 - 4 s_{W}^2}} c_{Z} - s_{Z}$ \\ \hline
            $C_3$ & $c_{Z} + \frac{\sqrt{3 - 4 s_{W}^2}}{3 - 2 s_{W}^2} s_{Z}$ \\ \hline
            $C_4$ & $\frac{\sqrt{3 - 4 s_{W}^2}}{3 - 2 s_{W}^2} c_{Z} - s_{Z}$ \\ \hline
            $C_5$ & $c_{Z} + \frac{3 - 2 s_{W}^2}{(3 - 4 s_{W}^2)^{3/2}} s_{Z}$ \\ \hline
            $C_6$ & $\frac{3 - 2 s_{W}^2}{(3 - 4 s_{W}^2)^{3/2}} c_{Z} - s_{Z}$ \\ \hline
        \end{tabular}
    \end{minipage}
    \caption{
    Coefficients $g_{B,f}$ for the interactions of neutral gauge bosons $B(=A,Z^1, Z^2)$ with SM fermions. For convenience, we write the coefficients in the form $g_{B,f} =g^{SM}_{Z,f}\times C_i$, where $g^{SM}_{Z,f}$ describes the SM interactions.}
    \label{tab:coeff}
\end{table}

Since $X^{q}$ is a non-diagonal matrix one sees from the above expressions that FCNCs will emerge. 
From the coefficients in Table \ref{tab:coeff}, we obtain  
\be
g_{Z^1,d}^{FV} = g_{Z^1,u}^{FV} = -\frac{g_L c_{W}}{\sqrt{3-4 s_{W}^2}}s_{Z}\quad\quad \mbox{and} \quad\quad g_{Z^2,d}^{FV} = g_{Z^2,u}^{FV} = -\frac{g_L  c_{W}}{\sqrt{3-4 s_{W}^2}}c_{Z}\,. 
\ee

As a result both the $Z^1$ and $Z^2$ gauge bosons will mediate FCNCs to quarks. 
Moreover, in the limit of interest, $w\gg v_{EW}$ ($\theta_Z\ll 1$), the flavor-violating interactions mediated by $Z^1 (\simeq Z)$ are suppressed.
As for the flavor-conserving $Z^1$ interactions in the same limit, they recover their SM form since $C_1, C_3,C_5 \to 1$ (see Table \ref{tab:coeff}), as expected~\cite{Singer:1980sw}.

\subsection{Brief summary on 3-3-1 constraints} 

Starting with the gauge sector, the mass and branching fraction measurements of the observed $Z^0$-boson (which corresponds to the $Z_1$-boson)~\cite{ParticleDataGroup:2024cfk} at the colliders, place severe constraints on the 3-3-1 breaking scale $w$.
Moreover, the existence of the $Z_2$-boson introduces new contributions to various low energy observables, such as electroweak precision ovservables \cite{Long:1999bny,Erler:2009jh}, $Z^0$-pole observables \cite{Babu:1997st}, muon properties (muon decay, $\mu \to 3e$ and muon g-2) \cite{Kelso:2014qka, Doff:2024hid}, FCNCs \cite{Cogollo:2012ek,Buras:2014yna},  polarized electron-nucleon scattering \cite{Barger:1997nf}, atomic parity violation \cite{Long:2000mzu,CarcamoHernandez:2005ka,Babu:1997st}, electron-quark contact interactions \cite{Barger:1997nf} etc. 
Experimental bounds on these observables restrict the  $Z_2$-boson mass and hence the 3-3-1 breaking scale $w$. A most stringent constraint comes from the invisible decay width of the $Z^0$-boson by assuming the number of neutrino generations equal to three \cite{Cogollo:2007qx,Dorsch:2024ddk}. 

 From collider searches, the current lower bounds on the masses of $W'$ and neutral ($Z_2$ and $X^0$) bosons set by the ATLAS and CMS collaborations are 6 TeV \cite {ATLAS:2019lsy,ParticleDataGroup:2024cfk} and 5.15 TeV \cite{CMS:2021ctt,ParticleDataGroup:2024cfk} (in the Sequential Standard Model). This pushes the lower limit of $w$ to $\sim$18 TeV and 13 TeV respectively. Although this bound is model-dependent and can be relaxed a bit depending on the available decay channels of the new gauge bosons, one does not expect it to vary drastically for 3-3-1 model.

Turning to the scalar sector, in order to make the scalar potential bounded from below one requires: $\lambda_{\{1,2,3,\sigma\}}>0$. Stringent constraints on the quartic coupling parameters $\lambda_i$, $\lambda_{ij}$, $\tilde\lambda_{ij}$ and $\lambda_{i\sigma}$ follow from the vacuum stability, perturbativity, perturbative unitarity and the mass-spectrum of scalar particles~\cite{Costantini:2020xrn,Dorsch:2024ddk,Sanchez-Vega:2018qje,Dias:2025oyb}. 

The mass of the pseudoscalar $A$ requires $\lambda_A>0$ (see Eq. \eqref{eq:mA}). To ensure one eigenvalue of the scalar mass matrix $M_s^2$ matches $m_h^2$, $\lambda_A$ must be small,  which is effectively achieved with $\sqrt{\lambda_A v_\sigma w} \sim \mathcal O(m_h)$. This choice keeps the pseudoscalar at the electroweak scale, while the other CP-even scalars acquire masses around $v_{EW}$, $w$ and $v_\sigma$, respectively \cite{Pinheiro:2022bcs} (note however, $\lambda_A$ cannot be too small either as LHC constraints from $h\to H_2H_2$ and $h\to AA$ require $(m_A,\; m_{H_2})> 62.5$ GeV (i.e. $m_h/2$) \cite{Coutinho:2024zyp}.)
Alternatively, a viable and safe choice is $\lambda_A\sim \mathcal O(w/v_\sigma)$, since it pushes the masses of $A$, $H_2$ and $H_3$ to the 3-3-1 scale~\cite{Pinheiro:2022bcs} leaving no light exotic Higgs boson (note, however, $\lambda_A$ cannot be too large either, as this would result in negative eigenvalues in the matrix $M_s^2$).

The last complex neutral scalar acquires a mass $m_{\varphi^0}\sim \mathcal O (w)$ (see Eq. \eqref{eq:mphi0}).
On the other hand, among the two charged scalars, $H_1^\pm$ acquires the mass $m_{H_1^\pm}\sim O(m_h)$ whereas $H_2^\pm$ is heavy, with mass $m_{H_2^\pm}\sim O(w)$, as shown in Eq. \eqref{eq:charged}.  Searches at LEP put a lower bound on the mass of any charged Higgs at $\sim$ 95 GeV \cite{Coutinho:2024zyp}.  

Finally, the heavy exotic quarks are expected to have masses larger than $\sim$ 1.5 TeV \cite{CMS:2022fck,ATLAS:2024gyc, Huitu:2024nap}, the limit changing slightly depending on the decay channels of the exotic quarks.

\section{Axion physics with photons}
\label{sec:axionprop}

\begin{figure}[!h]
\begin{center}
\includegraphics[width=14.5cm]{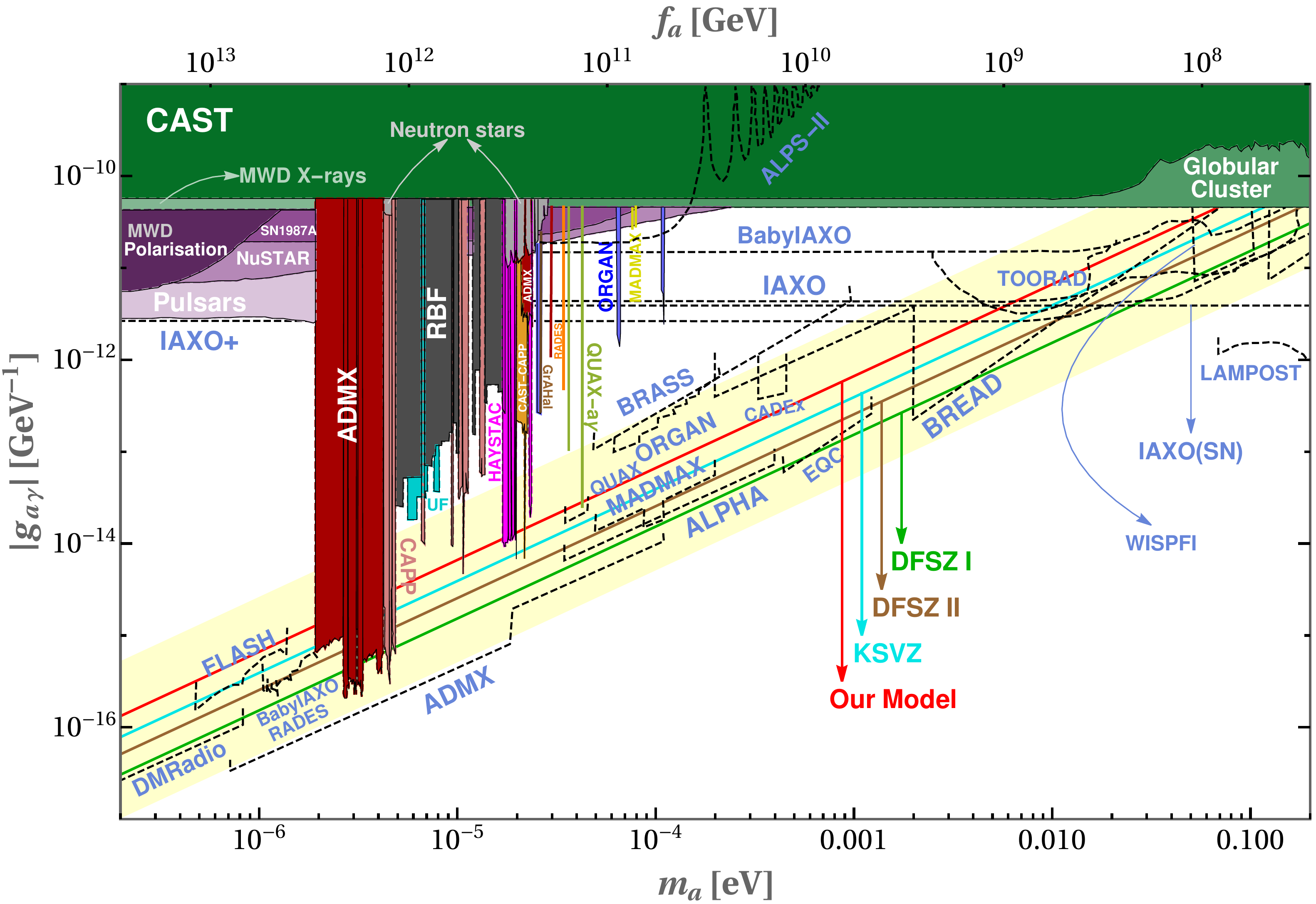} 
\caption{
Enhanced axion-to-photon coupling (in GeV$^{-1}$) versus axion mass and decay constant. Our model predictions are compared to those of the KSVZ, DFSZ I and DFSZ II schemes. Current experimental bounds and future projections~\cite{axionlim,axionpdg} are given (black dashed). See text for details.}
\label{plot:agg}
\end{center}
\end{figure}

The Peccei-Quinn symmetry has an associated non-vanishing $[\mathrm{SU(3)_c}]^2\times \mathrm{U(1)_{PQ}}$ anomaly coefficient 
\begin{eqnarray}\label{Cag}
 C_{ag}&=& \sum_{q} \left(x_{q_L}-x_{q_R}\right) = x_{\sigma}\,,
\end{eqnarray}
where the contributions from all quarks $q$ must be taken into account. 

The axion field, defined in Eq. (\ref{eq:axion}), can also be generically written as\,\cite{Srednicki:1985xd}
\begin{equation}\label{axion2}
 a = \frac{1}{f_{a}}\left[  v_1\,x_{1}\, a_1 +  v_2 \,x_{2}\, a_2 +  w \,x_{3}\, a_3 +  v_\sigma\, x_{\sigma}\,a_\sigma \right]\,.
\end{equation}
In the present model the PQ charges ($x$) are expressed in terms of two angles: 
\begin{eqnarray}
&\frac{x_{1}}{x_{\sigma}} =  - (\cos{\delta}\cos{\beta})^2,\quad\quad
\frac{x_{2}}{x_{\sigma}} = - (\cos{\delta}\sin{\beta})^2,\quad\quad
\frac{x_{3}}{x_{\sigma}} = - (\sin{\delta})^2,&\\
\label{angles}
&\text{where,} \quad 
\tan{\delta} = \frac{v_1v_2}{ w v_{EW}}\quad\quad\mbox{and} \quad\quad \tan{\beta} = \frac{v_1}{v_2}.&
\end{eqnarray}
With a domain wall number equal to unity, the axion decay constant is given by
\begin{eqnarray}\label{fa}
f_a = \sqrt{x_{\sigma}^2 v_\sigma^2+x_{1}^2 v_1^2+x_{2}^2 v_2^2+x_{3}^2 w^2}= x_{\sigma} \sqrt{ v_\sigma^2 + \frac{v_1^2 v_2^2 w^2}{ v_1^2 v_2^2 +v_{EW}^2 w^2} }\,,
\end{eqnarray}  
assuming that $x_{\sigma}>0$. Thus, in the limit $v_\sigma\gg w \gg v_1,v_2$, we have that $f_a\simeq v_\sigma$ so that $a\simeq a_{\sigma}$.

As usual, non-perturbative QCD effects generate the PQ-scale suppressed axion mass~\cite{Weinberg:1977ma,diCortona:2015ldu}
\begin{equation}
 m_a
 \simeq 5.7\left(\frac{10^{12}\,\mbox{GeV}}{f_a}\right) \mbox{$\mu$eV}\,,
\end{equation}
so that for $f_a\simeq v_\sigma$ in the range from $10^{9}-10^{12}\,\mbox{GeV}$, the axion becomes a cold dark matter candidate~\cite{Sikivie:2020zpn,DiLuzio:2020wdo} capable of providing the full 
relic dark matter density from Planck observations~\cite{Planck:2018vyg}.
 
We now turn to the axion-photon coupling, that depends on the $[\mathrm{U(1)_Q}]^2\times \mathrm{U(1)_{PQ}}$ anomaly coefficient, defined as 
\begin{equation}\label{cagamma}
  C_{a\gamma} = 2 \sum_{f}(x_{f_L}-x_{f_R})Q^2 =  -\frac{4}{3} x_{\sigma}\,,
\end{equation}
where the sum is over all the fermions $f$ in the theory. The resulting effective Lagrangian is 
\be
{\cal L}_{a\gamma\gamma}=-\frac{g_{a\gamma}}{4}a\,F_{\mu\nu}\tilde{F}^{\mu\nu},
\label{laxga}
\ee
with the axion-to-photon coupling 
\be
g_{a\gamma}=\frac{\alpha}{2\pi f_a}\left(\frac{C_{a\gamma}}{C_{ag}}-\frac{2}{3}\frac{4+z}{1+z}\right)\approx\frac{\alpha}{2\pi f_a}\left(-\frac{4}{3}-1.92\right) \,,
\label{gag}
\ee
where $\alpha$ is the fine-structure constant and $z = m_u/m_d\approx 0.596$ is the up- and down-quark mass ratio.

The different value for the $C_{a\gamma}/C_{ag}$ ratio -- often also referred to as $E/N$ -- when compared to the more conventional axion schemes is related to the intrinsic flavor structure of 3-3-1 scenarios~\cite{Singer:1980sw,Valle:1983dk,Frampton:1992wt,Pisano:1991ee}. The latter leads to an enhanced $g_{a\gamma}$ coupling, as well as to a ``flavored'' axion, in the absence of an imposed flavor symmetry, see Sec.\ref{sec:charged-leptons}. 

In Fig. \ref{plot:agg} we compare our $|g_{a\gamma}|$ coupling as a function of the axion mass $m_a$ with the results from other well-known models.  Our prediction ($C_{a\gamma}/C_{ag}=-4/3$) is indicated by a red line. The other models KSVZ, DFSZ I and DFSZ II which predict the value of $C_{a\gamma}/C_{ag}$ as $ 0$, $8/3$ and $2/3$ are depicted by cyan, green and brown lines, respectively. The pale yellow band indicates the theoretical region for any QCD-axion model with $5/3\leq C_{a\gamma}/C_{ag} \leq 44/3$. Current limits from various experiments are shown in different solid lines and shaded colors, while the future projections are indicated in black dashed lines. For details of all these experimental searches, see ~\cite{axionlim,axionpdg}. In our model Globular Cluster limits rule out axion masses larger than 70 meV. On the other hand, ADMX and CAPP combined rules out the axion mass range from 1.95 $\mu$eV to 4.79 $\mu$eV, while HAYSTAC does the same for the mass range of 17.18 $\mu$eV to 19.36 $\mu$eV. Apart from these excluded ranges, we note that CAPP, CAST-CAPP and QUAX-$a\gamma$ rule out a few specific axion masses. Thus, our model is currently allowed for axion masses $m_a\leq$ 70 meV, excluding the ranges mentioned above. However, improved experimental sensitivities could conceivably rule out or validate our model at various mass ranges. We stress again its enhanced coupling parameter $|g_{a\gamma}|$  when compared to KSVZ and DFSZ models.

\section{Axion couplings to matter} 

We can now obtain the axion couplings to fermions in our scheme. They follow from the Yukawa interactions in Eqs. (\ref{lagYl}) and (\ref{lagYq}) and the axion profile given in Eq. (\ref{axion2}). The generic interaction Lagrangean is
\begin{equation}\label{axionFCNC}
     \mathcal{L}_{af} =i a\, \overline{f_i}\left[\left(g_{af}^V\right)_{ij}- \left(g_{af}^A\right)_{ij}\gamma^5\right]f_j.
\end{equation}
For charged leptons in the mass basis, $f = e$, the relevant coefficients are given as 
\begin{equation}
\left(g_{ae}^V\right)_{ij} = 0\quad\quad \mbox{and} \quad\quad \left(g_{ae}^A\right)_{ij} = -\frac{m_i\, \delta_{ij}}{f_a}\,\cos^2{\delta}\,\sin^2{\beta}\,.
\end{equation}
Therefore, the axion couples diagonally to the charged-lepton flavors. 

Turning to the neutral fermions, also in the mass basis, we find that, in addition to the flavor-diagonal terms, there are non-diagonal axion interactions. These involve fields with different $\mathrm{U(1)_{PQ}}$ charges and follow from the Dirac seesaw mechanism: 
\begin{eqnarray}
(g_{a{N}}^V)_{ij} &=& \frac{m^{N}_i-m^{N}_j}{2 f_a}\left[(1+\cos^2\delta\cos^2\beta)\times X^{N_L}_{ij} - (1+ \sin^2\delta)\times X^{N_R}_{ij}\right],\\
(g_{aN}^A)_{ij} &=& \frac{m^{N}_i+m^{N}_j}{2 f_a}\left[(\cos^2\delta\sin^2\beta-2)\times \delta_{ij}+(1+\cos^2\delta\cos^2\beta)\times X^{N_L}_{ij} + (1+ \sin^2\delta)\times X^{N_R}_{ij}\right],\nonumber
\end{eqnarray}
\begin{equation}
 \text{with } \quad X^{N_{L,R}}_{ij} =\left[(U^N_{L,R})^\dagger\mbox{diag}\left(0_{3\times 3},I_{3\times 3}\right)U^N_{L,R}\right]_{ij}~.
\end{equation}
Here the entries $m_i^N$ represent the masses of both active neutrinos and heavy neutral leptons, and $i,j = 1,\cdots,6$. \\ 

Turning finally to the axion couplings to mass-eigenstate quarks, we find both flavor-conserving as well as flavor-violating interactions
\begin{eqnarray}
&\left(g_{aq}^V\right)_{ij}=\frac{m^{q}_i-m^{q}_j}{2 f_a}\cos^2\delta\times X^{q}_{ij}\,,&\nonumber\\
&\left(g_{a{u}}^A\right)_{ij} = \frac{m^{u}_i+m^{u}_j}{2 f_a}\cos^2\delta\left[\sin^2\beta \times \delta_{ij} + X^{u}_{ij}\right]\quad \text{and} \quad \left(g_{a{d}}^A\right)_{ij} = \frac{m^{d}_i+m^{d}_j}{2 f_a}\cos^2\delta\left[\cos^2\beta  \times \delta_{ij} + X^{d}_{ij}\right]\,,&
\end{eqnarray}
where with $q = u, d$, with $X^{q}_{ij}$ is defined in Eq. (\ref{eq:Xq}). 
Once again, we emphasize that the non-standard embedding of quark families in $\mathrm{SU(3)_L}$ representations leads to flavor-changing NC quark interactions. Here, these interactions involve the axion and are encoded in the matrices $X_{ij}^{u,d}\not\propto \delta_{ij}$. 
Finally, the axion couplings to the exotic quarks, $D_\alpha$ and $U_3$ are diagonal and described by $g_{aD} =\mbox{diag}(m_{D_1},m_{D_2})\sin^2{\delta}/{f_a}$ and $g_{aU} =-m_{U}\sin^2{\delta}/{f_a}$. \\[-.2cm]

In the remainder of this section we discuss the physics associated to flavor-conserving axion couplings to matter.

\subsection{Axion-to-electron coupling}
\label{sec:elec-axion}

The Lagrangian describing the axion interaction to electrons is given by:
\begin{equation}
    \mathcal L_{ae}=-ig_{ae} \,a\, \bar e \gamma^5 e \quad \text{where} \quad g_{ae}=\frac{m_e}{f_a} c_{ae} \quad \text{with} \quad c_{ae}=\frac{x_{eL}-x_{eR}}{x_\sigma}=-\cos^2\delta \sin^2\beta.
\end{equation}
Since current limits on 3-3-1 gauge bosons require $w\gsim20$ TeV, we have $\cos^2\delta\approx1$ implying $g_{ae}\approx-\frac{m_e}{f_a} \sin^2\beta$. This situation looks similar to the DFSZ scenario, except for the fact that the coupling is three times higher. The number of domain-walls is responsible for this enhancement.  This should be clear, given that our model approaches a two-Higgs-doublet effective axion model in the low energy limit.

\begin{figure}
    \centering
    \includegraphics[width=14.5cm]{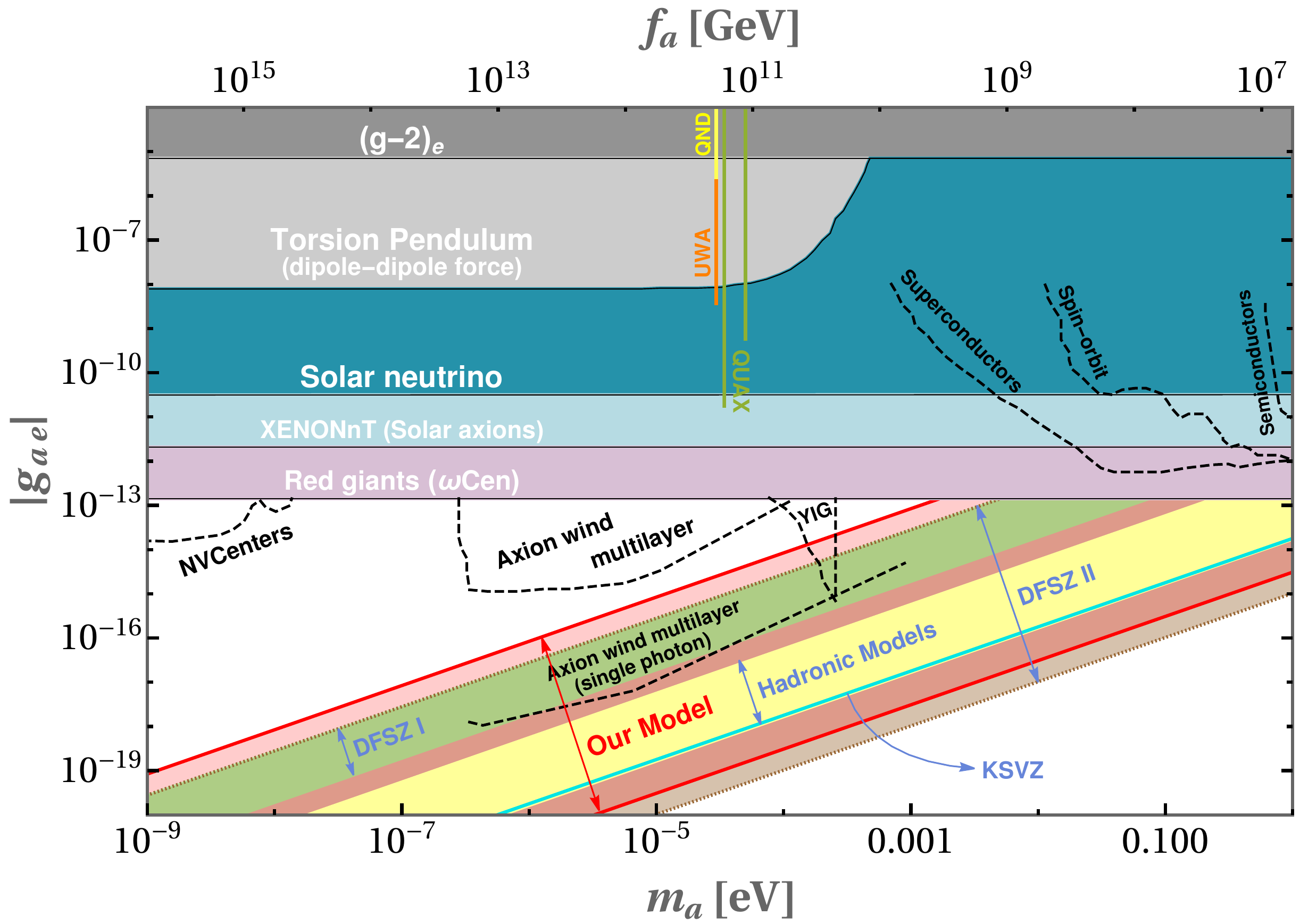}
    \caption{
    Axion-to-electron coupling versus axion mass and decay constant. Our model predictions are compared to those of the KSVZ, DFSZ I and DFSZ II schemes. Current experimental bounds and future projections~\cite{axionlim,axionpdg} are also presented (black dashed). See text for details.}
    \label{fig:aee}
\end{figure}

In Fig. \ref{fig:aee} we compare axion-electron coupling with theoretical predictions from other models, as well as current experimental bounds and future projections \cite{axionlim,axionpdg,Langhoff:2022bij,Gondolo:2008dd,DarkSide:2022knj,GERDA:2020emj,PandaX:2017ock,SuperCDMS:2019jxx,XENON:2020rca,XENON:2022ltv,VanTilburg:2020jvl,Capozzi:2020cbu,QUAX:2020adt,Ferreira:2022egk}. 
In DFSZ I and DFSZ II scenarios, the axion-electron coupling at tree-level becomes~\cite{DiLuzio:2020wdo,Srednicki:1985xd}:
\begin{equation}
    c_{ae}^{\text{DFSZ I}} = \frac{1}{3}\sin^2 \beta^{\text{DFSZ}} \quad \text {and} \quad c_{ae}^{\text{DFSZ II}} = -\frac{1}{3}\cos^2 \beta^{\text{DFSZ}} \quad \text{with} \quad \tan \beta^{\text{DFSZ}}=\frac{v_u}{v_d}.
\end{equation}
However, perturbative unitarity of the Yukawa couplings of the SM fermions requires: $0.25\leq \tan \beta^{\text{DFSZ}}\leq 170$ \cite{DiLuzio:2020wdo} (for both the DFSZ scenarios) which implies: $0.0196\leq|c_{ae}^{\text{DFSZ I}}|\leq0.333$ and $1.1 \times 10^{-5}\leq|c_{ae}^{\text{DFSZ II}}|\leq0.314$. These two cases are shown by the green and brown bands in Fig. \ref{fig:aee}. It is evident from Eqs. \eqref{lagYl} and \eqref{lagYq} that our scenario resembles the DFSZ II model regarding couplings of scalars to the first and second generations of fermions with $\Phi_2^*\equiv H_u$ and $\Phi_1^*\equiv H_d$.  Thus, the angle $\beta$, defined in Eq. \eqref{lagYl}, has a phase-shift of $\pi/2$ compared to the corresponding DFSZ angle $\beta^{\text{DFSZ}}$, i.e. $\tan\beta\equiv \cot \beta^{\text{DFSZ}}$, so that the electron-axion coupling in our case is bounded by: $3.46\times 10^{-5}\leq c_{ae}\leq0.941$ (red band). For the KSVZ scenario ($E/N=0$) and other hadronic axion models ($5/3\leq E/N \leq 44/3$) the electron-axion coupling at tree-level vanishes, but is generated radiatively at one-loop:  $c_{ae}^{\text{1-loop}}= (3\alpha_{em}^2/4\pi^2)[(E/N)\log(f_a/m_e)-1.92\log(\Lambda_\chi/m_e)]$ with $\Lambda_\chi\simeq 1$ GeV being the chiral symmetry breaking scale. These are depicted with the cyan line and the yellow band in Fig. \ref{fig:aee}. 

Note that the KSVZ model predicts a line for the electron-axion coupling, while DFSZ I predicts a narrow band. In contrast, both the DFSZ II and our scenario predict different broad bands for the electron-axion coupling. A difference between them emerges from the smaller domain-wall-number for the case of the revamped axion. This translates into substantial differences in the allowed axion mass ranges.
As seen from Fig. \ref{fig:aee}, red giants provide the strongest bound on the electron-axion coupling $|g_{ae}|$. Below 1.58 meV axion mass, astrophysics does not constrain our model at all. It seems unlikely that future improvements will be able to severely restrict the $|g_{ae}|$ sensitivity of our revamped axion scenario.

\subsection{Axion-to-hadron coupling}
\label{sec:had-axion}

\begin{figure}
    \centering
    \includegraphics[width=14.5cm]{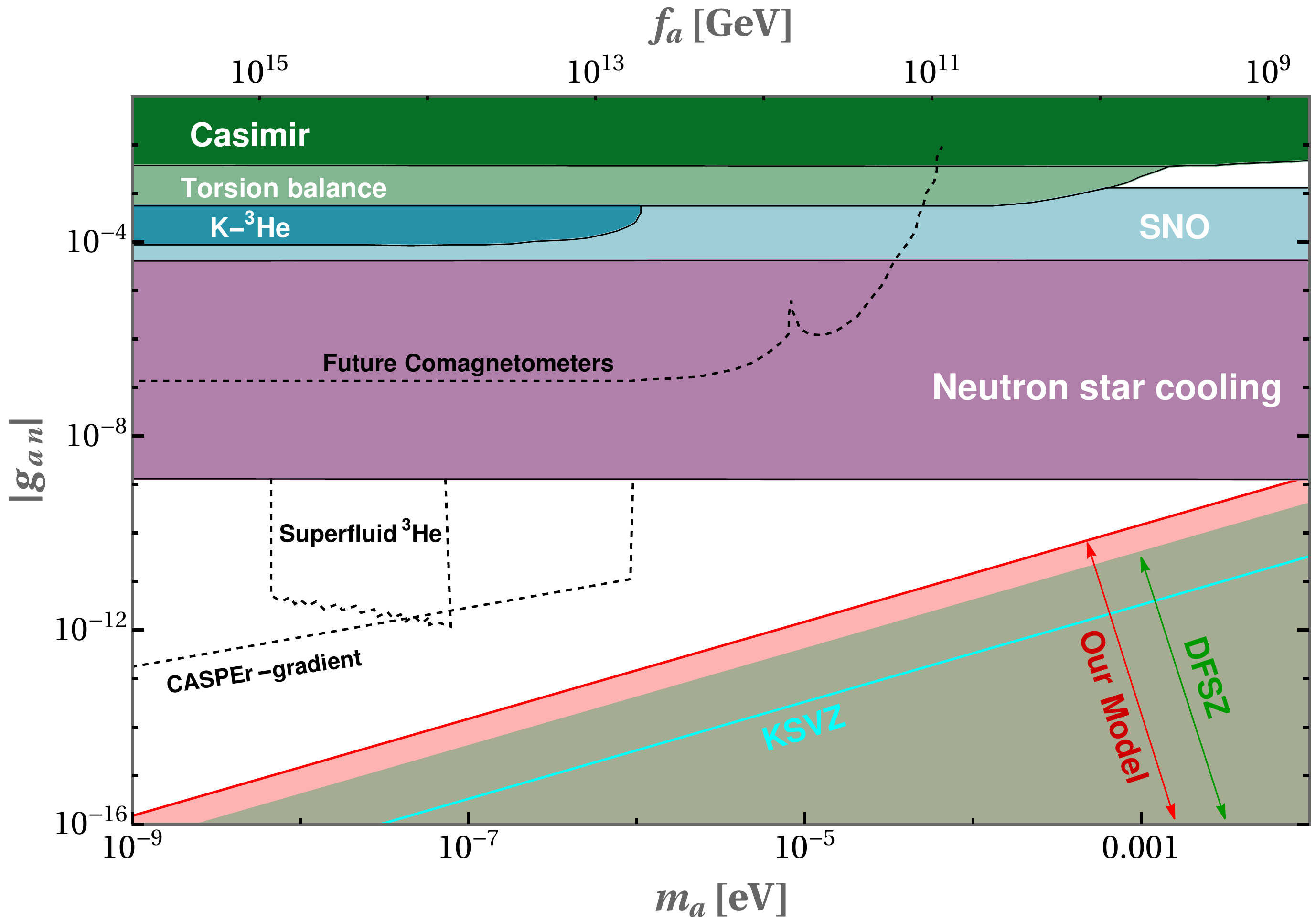}
    \caption{ 
    Axion-to-neutron coupling as function of axion mass (in eV). Our model is compared to KSVZ, DFSZ (both I and II) schemes. Current experimental bounds and future projections (black dashed) \cite{axionlim,axionpdg} are also indicated.}
    \label{fig:an}
\end{figure}

The flavour-conserving Lagrangian describing the interaction of the axion to
quarks is given by:
\begin{eqnarray}
\label{lag:qa}
     &\mathcal L_{aq_j^{}}=-ig_{aq_j^{}} \,a\, \bar q_j \gamma^5 q_j \quad \text{where} \quad g_{aq_j^{}}=\frac{m^q_j}{f_a}\, c_{aq_j^{}}&\nonumber\\ &\text{with} \quad c_{au_j^{}}=\cos^2\delta\,[\sin^2\beta+X^u_{jj}] \quad \text{and} \quad c_{ad_j^{}}=\cos^2\delta\,[\cos^2\beta+X^d_{jj}].&
\end{eqnarray}
The matrix $X^{u,d}_{ij}$ is defined in Eq. (\ref{eq:Xq}). As already mentioned, the scale of $w$ makes $\delta$ very small, so that: $c_{au_j^{}}\approx\sin^2\beta+X^q_{jj}$ and $c_{ad_j^{}}\approx\cos^2\beta+X^q_{jj}$, see Eq.~\ref{angles}.
On the other hand, the Lagrangian for axion-pion and axion-nucleon interactions can be expressed as \cite{DiLuzio:2020wdo,Chang:1993gm}:
\begin{eqnarray}
     &\mathcal L_{a\pi}=\frac{c_{a\pi}}{f_a f_\pi} \;\partial_\mu a \;(2\partial^\mu \pi^0 \pi^+\pi^- - \pi^0 \partial^\mu \pi^+\pi^- - \pi^0 \pi^+ \partial^\mu \pi^-)&\\
     & \text{and}\quad \mathcal L_{a\mathcal N}=-ig_{a\mathcal N} \,a\, \overline {\mathcal N} \gamma^5 \mathcal N \quad \text{with} \quad g_{a\mathcal N}=\frac{m_\mathcal N}{f_a} c_{a\mathcal N}^{} \quad \text{where}\quad \mathcal{N} \in \{p,n\}\;.&
\end{eqnarray}
Within the non-relativistic effective Lagrangian approach for the axion-nucleon interaction at leading order, the chiral expansion for the axion-pion coupling and axion-nucleon couplings (including running effects) can be expressed as 
\cite{GrillidiCortona:2015jxo}:
\begin{align}
&c_{a\pi}=0.12 +\frac{1}{3} \left(c_{ad}-c_{au}\right),\nonumber\\
&c_{ap}=-0.47+0.88\,c_{au}-0.39\,c_{ad}-c_{a,\,sea}\,,\nonumber\\
&c_{an}=-0.02-0.39\,c_{au}+0.88\,c_{ad}-c_{a,\,sea}\,,\nonumber\\
&c_{a,\,sea}=0.038\,c_{as} +0.012\,c_{ac} +0.009\,c_{ab} +0.0035\,c_{at}\,.
\end{align}
Axion-nucleon couplings with the chiral Lagrangian at the next-to-next-leading order level can be found in Ref. \cite{Vonk:2020zfh}. 

\begin{figure}
    \centering
    \includegraphics[width=14.5cm]{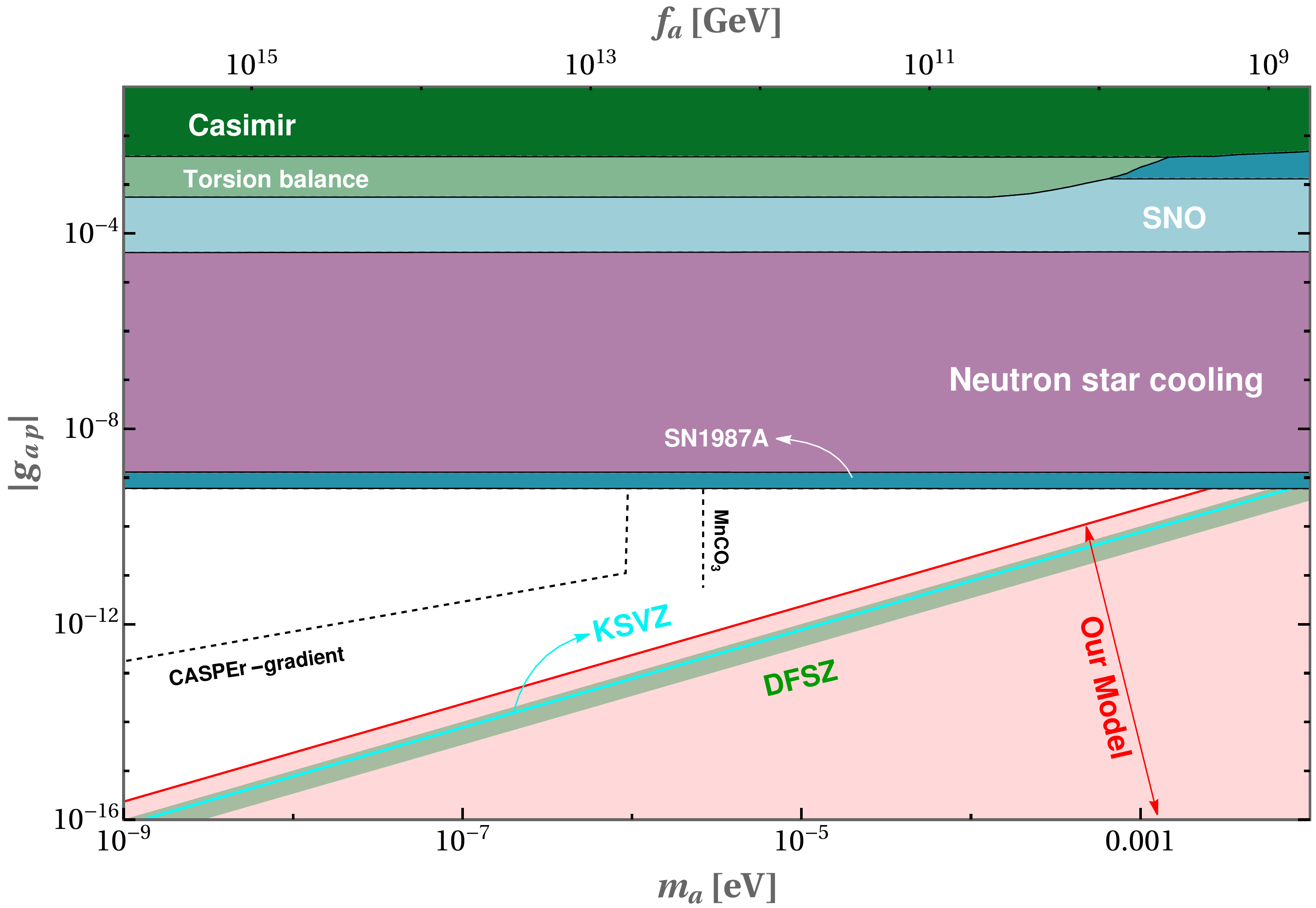}
    \caption{ 
    Axion-to-proton coupling versus the axion mass (in eV). Our model is compared to KSVZ, DFSZ (both I and II) schemes. Current experimental bounds and future projections (black dashed) \cite{axionlim,axionpdg} are also indicated.}
    \label{fig:ap}
\end{figure}

Within the KSVZ scenario, $c_{aq_j^{}}=0$ for all SM quark flavours at the tree level, leading to:
\begin{equation}
    c_{a\pi}^{\text{KSVZ}}=0.12\,, \quad 
    c_{ap}^{\text{KSVZ}}=-0.47 \quad \text{and} \quad c_{an}^{\text{KSVZ}}=-0.02\,.
\end{equation} 
In contrast, both DFSZ scenarios give: $c_{au_j^{}}=\frac{1}{3}\cos^2 \beta^{\text{DFSZ}}$ and  $c_{ad_j^{}}=\frac{1}{3}\sin^2 \beta^{\text{DFSZ}}$ at the tree level.  
Requiring perturbative unitarity of the Yukawa couplings of the SM fermions leads to the constraint on $\beta^{\text{DFSZ}}$ (i.e. $\beta^{\text{DFSZ}}\leq 170$ \cite{DiLuzio:2020wdo}) and therefore:
\begin{equation}
    0.022\leq c_{a\pi}^{\text{DFSZ}}\leq 0.231\,, \quad 
    -0.616\leq c_{ap}^{\text{DFSZ}} \leq -0.207  \quad \text{and} \quad -0.131 \leq c_{an}^{\text{DFSZ}}=0.258\,.
\end{equation}
Using similar perturbativity constraint in our scenario, we find:
\begin{equation}
    -0.249\leq c_{a\pi}\leq 0.528\,,\quad -1.425 \leq c_{ap} \leq 0.396 \quad \text{and} \quad -0.874 \leq c_{an}\leq 0.894.
\end{equation}

In Figs. \ref{fig:an} and \ref{fig:ap}, we present the axion-nucleon couplings  $|g_{an}|$ and $|g_{ap}|$ as functions of the axion mass $m_a$. We compare our theoretical predictions for the ranges of these couplings with those of the KSVZ and DFSZ models. We also present the current experimental bounds and future projections from different experiments \cite{Abel:2017rtm,JEDI:2022hxa,Adelberger:2006dh,Mostepanenko:2020lqe,Wu:2019exd,Garcon:2019inh,Wei:2023rzs,Jiang:2021dby,Bloch:2022kjm,Bloch:2019lcy,Vasilakis:2008yn,Lee:2022vvb,Buschmann:2021juv,Bhusal:2020bvx,Lella:2023bfb,Chigusa:2023hmz}, see \cite{axionlim,axionpdg}. One sees from Fig. \ref{fig:an} that, for any value of $m_a$, our model allows a somewhat bigger range of variation for $|g_{an}|$ than the DFSZ scenario. However, a striking difference arises in Fig. \ref{fig:ap}. Although DFSZ predicts a band with very small width for $|g_{ap}|$, our model allows a much wider range of $|g_{ap}|$ for any particular value of $m_a$. This occurs because of presence of the $X_{jj}^{u,d}$ term in quark-axion coupling (see Eq. \eqref{lag:qa}). Unfortunately, in contrast to the axion-to-photon coupling, probing the coupling to nucleons seems to pose a very demanding task.

\section{Flavor violation with axions } 
\label{sec:charged-leptons}

As already mentioned, our model generates flavor-violating axion-quark couplings at tree-level.  These interactions can force a heavier pseudoscalar meson $P_i$ to decay into a lighter one ($P_j$) by emitting the axion. Defining $P_i\equiv \bar q_i q$, the branching fractions for these processes are given generically as \cite{Bjorkeroth:2018dzu, MartinCamalich:2020dfe}:
\begin{equation}
\label{eq:meson-axion}
    \mathcal{B} (P_i\to P_j\,a)=\frac{m_i^3\,|V^q_{ij}|^2}{16\pi \,f_a^2\,\Gamma_i }  \left(1-\frac{m_j^2}{m_i^2}\right)^3 |f_+^{ij}(0)|^2 \quad \text{with} \quad  V^q_{ij}=\frac{1}{2}X_{ij}^q\cos^2\delta
\end{equation} 
where, $m_i$ and $\Gamma_i$ denote the mass and decay-width of the $P_i$ meson, $f_+^{ij}(0)$ is the form-factor related to the  $P_i\to P_j$ transition at $q^2\to 0$, and the factor $V_{ij}^q$ is related to the quark-axion coupling. Lattice results for the form factors $f_+^{ij}(0)$ are presented in Tab. \ref{tab:form_factor}.
\begin{table}[ht!]
   \begin{tabular}{|c|c|c|c|c|c|c|c|c|}
    \hline
        ~Form-factor~ & ~$f_+^{K\pi}(0)$~  & $f_+^{BK}(0)$ & $f_+^{B\pi}(0)$ & $f_+^{B_s K}(0)$ & $f_+^{D\pi}(0)$ & $f_+^{D_s K}(0)$ \\
        \hline
         Value & 0.97 & ~0.32(6)(6)~ & ~0.27(7)(5)~  &
         ~0.23(5)(4)~ & ~0.74(6)(4)~ &
         ~0.68(4)(3)~ \\
         \hline
    \end{tabular}
    \caption{
    Lattice results for the form factors $f_+^{ij}(0)$ corresponding to the $P_i\to P_j$ transitions \cite{Al-Haydari:2009kal,FlavourLatticeAveragingGroupFLAG:2024oxs}. The numbers quoted in the parentheses indicate the statistical and systematic errors respectively.}
    \label{tab:form_factor}
\end{table} 

Several experimental searches for the decays $P_i\to P_j\,a$ have already been performed \cite{ParticleDataGroup:2024cfk}. Since our pseudoscalar $a$ will act as a QCD axion, we restrict ourselves to the searches with $m_a\ll 1$ GeV. 
The relevant experimental searches are compiled in Tab. \ref{tab:expt_meson_axion}. One sees that the most stringent constraint on such processes come from $K^+\to \pi^+ a$ decay, whereas there is no search in the charmed meson ($D$) sector. 
\renewcommand{\arraystretch}{1.3}
\begin{table}[ht!] 
    \centering
    \begin{tabular}{|c|c|c|c|}
    \hline
     \multirow{2}{*}{Decay} & Quark & Branching & \multirow{2}{*}{Experiment} \\
     & transition & fraction & \\
     \hline
       \multirow{5}{*}{$K^+\to \pi^+ a$} & \multirow{5}{*}{$\bar s \to \bar d\,a$} & $<0.45\times10^{-10}$ & E787 (2002) \cite{E787:2002qfb,E787:2004ovg} \\
       & & $<0.59\times10^{-10}$ & E787 (2001) \cite{E787:2001urh}\\
       & & $<0.73 \times 10^{-10}$ & E787 + E949 (2007) \cite{E949:2007xyy} \\
       & & $<1.2\times10^{-10}$ & E949 (2009) \cite{BNL-E949:2009dza}\\
       & & $<2.0\times10^{-10}$ & NA62 (2020) \cite{NA62:2020xlg}\\
       \hline
    \multirow{2}{*}{$K_L^0\to \pi^0 a$}  & \multirow{2}{*}{$\bar s \to \bar d\,a$} & $<2.25 \times 10^{-9}$ & KOTO (2018)
    \cite{KOTO:2018dsc} \\
    & & $<5.0 \times 10^{-8}$& KOTO (2016) \cite{KOTO:2016vwr}\\
    \hline
    $B^\pm\to K^\pm a$ & $\bpar{b}\to  \bpar s \, a$ & $<4.9\times 10^{-5}$ & \multirow{2}{*}{CLOE (2001) \cite{CLEO:2001acz}}\\
    $B^0\to K^0_S a$ & $\bar{b}\to  \bar s \, a$ & $<5.3\times 10^{-5}$ & \\
    \hline
    $B^\pm\to \pi^\pm a$ & $\bpar{b}\to  \bpar d \, a$ & $<4.9\times 10^{-5}$ & CLOE (2001) \cite{CLEO:2001acz}\\
    $B_s^0\to \overline K^0 a$ & $\bar{b}\to  \bar d \, a$ & $<1$ & \\
    \hline
    $D^\pm\to \pi^\pm a$ & $\bpar{c}\to  \bpar u \, a$ & $<1$ & \\
    $D^0\to \pi^0 a$ & $c \to u \, a$ & $<1$ & \\
    $D_s^\pm\to K^\pm a$ & $\bpar{c}\to  \bpar u \, a$ & $<1$ & \\
    \hline
    \end{tabular}
    \caption{
    Relevant axion-emitting two-body decays of pseudoscalar mesons and their experimental limits.
    }
    \label{tab:expt_meson_axion}
\end{table}

\begin{figure}[!ht]
\begin{center}
\includegraphics[scale=0.46]{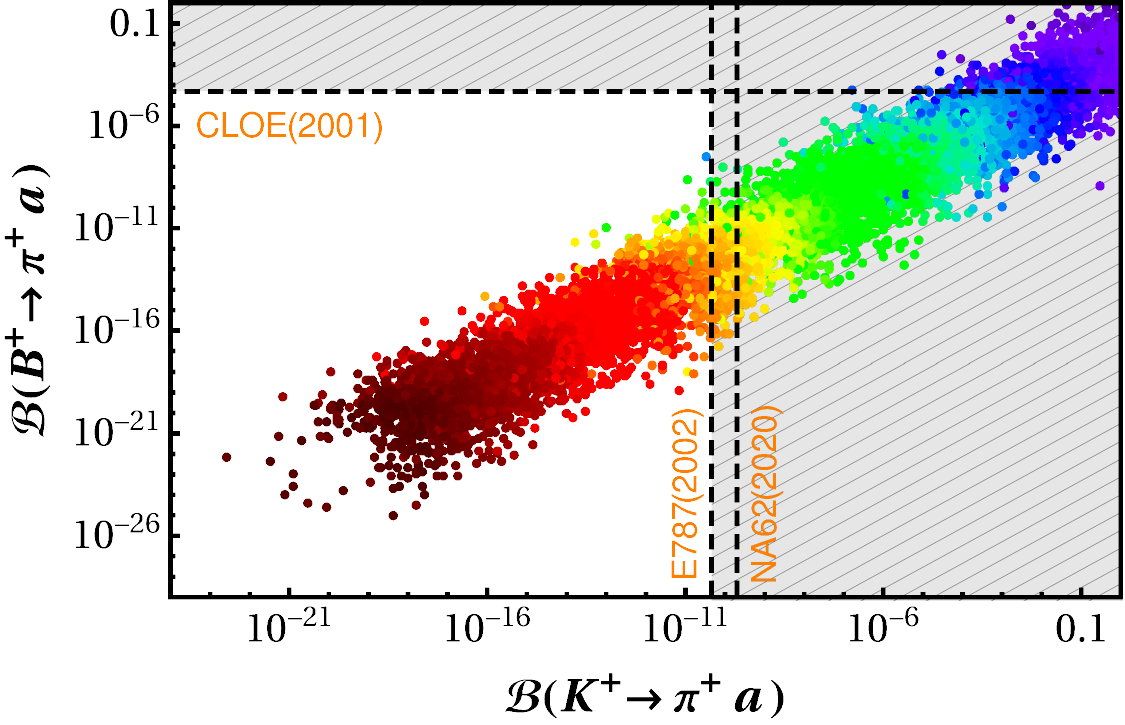}
\hfil
\includegraphics[scale=0.46]{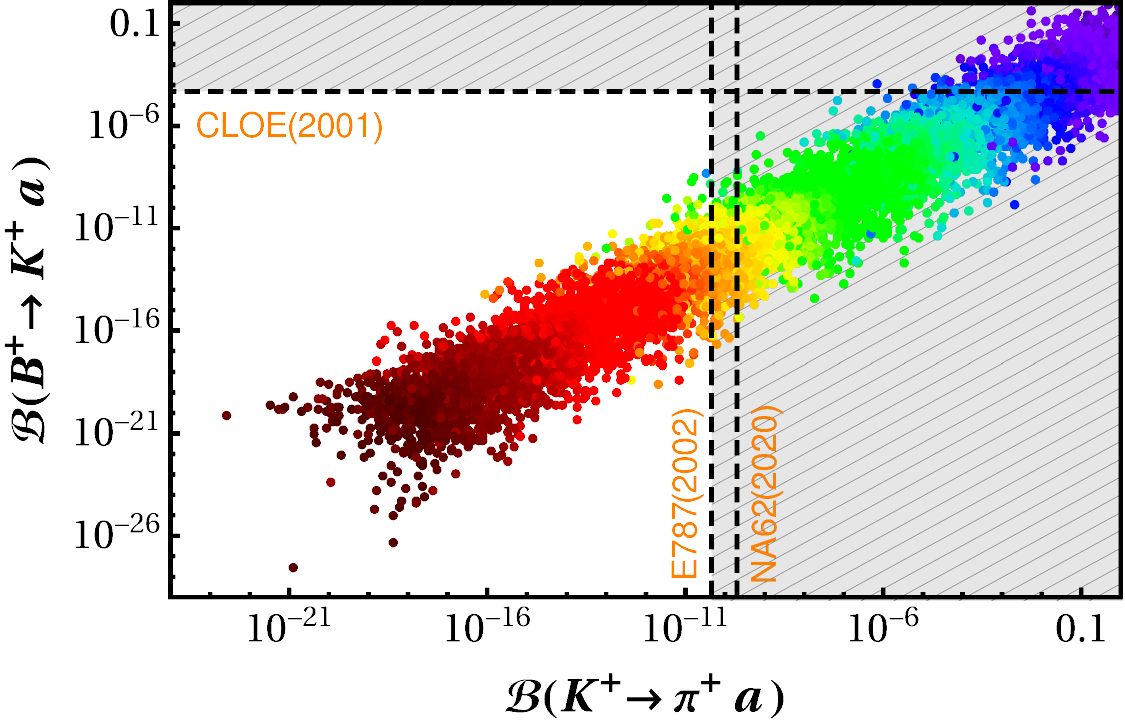}

\vspace*{5mm}
\includegraphics[scale=0.46]{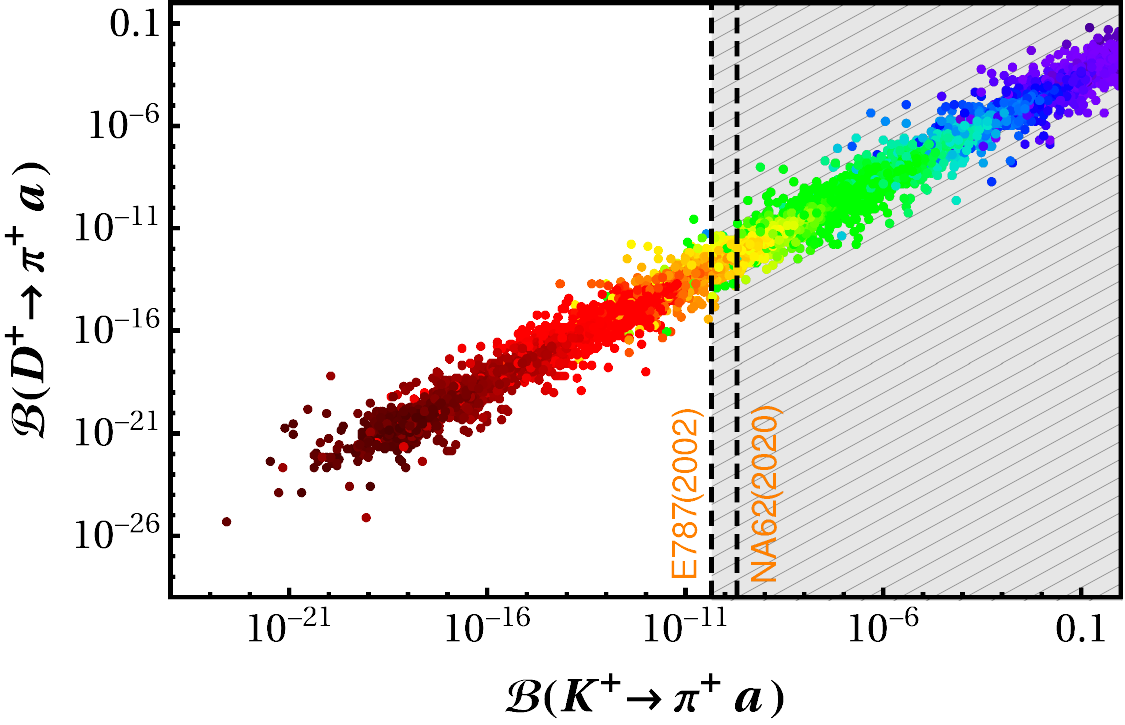}
\hspace{1cm} 
\includegraphics[trim={0 -2.3cm 0 0},clip,scale=0.35]{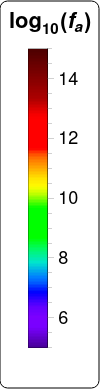}
\caption{
Correlation of $\mathcal B (B^\pm\to \pi^\pm a)$, $\mathcal B (B^\pm\to K^\pm a)$ and $\mathcal B (D^\pm\to \pi^\pm a)$ with $\mathcal{B}(K^+\to \pi^+ a)$ which translate to the correlation between $b\to d\,a$, $b\to s\,a$ and $c\to u\, a$ transitions with $s\to d \, a$ at the quark level.  The color of the points indicates the PQ-symmetry breaking scale $v_\sigma$ or $f_a$.}
\label{plot:brs}
\end{center}
\end{figure}
\begin{figure}[!h]
    \begin{center}
\includegraphics[scale=0.6]{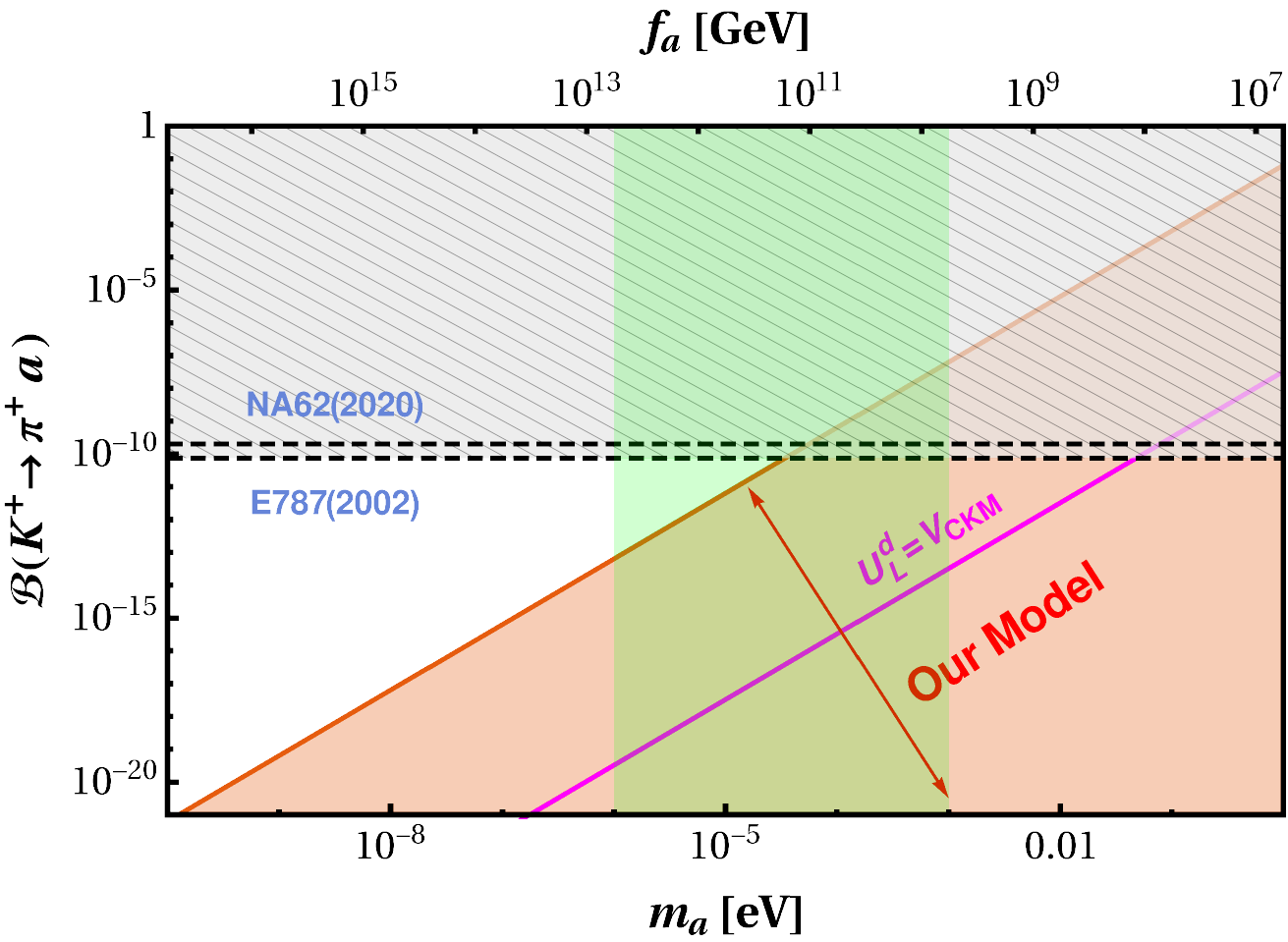}
\caption{
Variation of $\mathcal B(K^+\to \pi^+ a)$ with $m_a$. The black dashed lines represent the current experimental bounds on $\mathcal B(K^+\to \pi^+ a)$ from NA62 and E787.  The magenta line indicates $U_L^u=I$, i.e. $U_L^d=V_{CKM}$, whereas the red-shaded region is the allowed range of $\mathcal B(K^+\to \pi^+ a)$ in our model.}
\label{plot:KpiaVsfa}
\end{center}
\end{figure}

One sees from Eq.~\eqref{eq:meson-axion} that the branching fraction for the axion-emitting flavor-violating meson decay depends on the model parameters through the factor $|V_{ij}^q/f_a|^2$. This factor contains the PQ-symmetry breaking scale $f_a$ or $v_\sigma$, the angle $\delta$ (given by Eq. \eqref{angles}) and the matrix $X^q$ which is defined in terms of the up- and down-type left-handed mixing matrices $U_L^q$ in Eq. \eqref{eq:Xq}. Since, ${v_1^2+v_2^2}=v_{EW}^2$, the value of $\tan \delta$ is always bounded from above, $\displaystyle \tan\delta = \left(\frac{v_1 v_2}{w\, v_{EW}}\right)\leq \left(\frac{v_{EW}}{2\,w}\right)$. Therefore, taking $w\gsim 20\,{\rm TeV}\gg v_{EW}$ we restrict the angle $\delta$ to very small value, i.e. $\delta\lsim 0.006$, which effectively makes $\cos^2\delta\approx 1$. 
On the other hand, the unitary matrices $U^u_L$ and $U^d_L$ are related to each other through CKM-matrix (see Eq. \eqref{eq:CKM}) so we take only one of them as an independent object. We take $U_L^u$ as the independent unitary matrix, and generate it randomly. 
Then $U^d_L$ is automatically generated through Eq. \eqref{eq:CKM}. Hence the branching fractions for the  $P_i\to P_j\,a$ decays will vary depending on the matrix $U_L^u$ and $f_a$. The correlations between $\mathcal B (B^\pm\to \pi^\pm a)$, $\mathcal B (B^\pm\to K^\pm a)$ and $\mathcal B (D^\pm\to \pi^\pm a)$ decay branching fractions with $\mathcal{B}(K^+\to \pi^+ a)$ 
are shown in Fig.~\ref{plot:brs}.
The different colors in the scatter-plot points indicate different values for the axion decay constant $f_a$, on a logarithmic-scale. One should notice from Tab. \ref{tab:expt_meson_axion} that any particular quark-level transition can be constrained from appropriate meson decay channels. The branching fractions for the channels with the same quark level transition are proportional to each other, as they differ only in the kinematics and the form-factor. Hence, for each quark-transition we have only considered the mode providing strongest constraint. 
\footnote{Note also that
one should not find any difference between $q_i\to q_j a$ and $\bar q_i\to \bar q_j a$ transitions from the meson decays.}
Note that the
limits on
$\mathcal B (B^\pm\to \pi^\pm a)$, $\mathcal B (B^\pm\to K^\pm a)$ are very weak, while $D^\pm\to \pi^\pm a$ has not been studied.
However, as seen from the plot, within our model the predicted decay branching ratio correlations, together with the strong limits on $\mathcal{B}(K^+\to \pi^+ a)$, suggest much stronger constraints
on $\mathcal B (B^\pm\to \pi^\pm a)$ and $\mathcal B (B^\pm\to K^\pm a)$ and also an upper bound on $\mathcal B (D^\pm\to \pi^\pm a)$.

Now we turn to the most constraining flavour-violating axion emitting meson decay. 
In Fig. \ref{plot:KpiaVsfa}, we present the variation of the $K^+\to \pi^+ a$ decay branching fraction with the axion mass $m_a$. The value of the axion decay constant is also indicated. The red area indicates our allowed region, which depends on the choice of the left-handed-quark-mixing matrix $U^u_L$. As as example, we  present the magenta line which indicates our model prediction if $U^u_L$ is chosen to be identity matrix, so that $U_d^L=V_{CKM}$.  For axion masses below 31.6 $\mu$ev there are no restrictions on the model from the $K^+\to \pi^+ a$ decay.
To guide the eye, we have also included a greenish vertical band indicating roughly the region for cold-dark-matter axion~\cite{Planck:2018vyg}.

\section{Conclusions}
\label{sec:Conclusions}

We have investigated the phenomenology of an interesting
3-3-1 extension of the Standard Model proposed in Ref.~\cite{Dias:2020kbj}, with special emphasis on the physics
associated to the axion.
In our setup the smallness of neutrino masses, the solution of the strong CP problem as well as the nature of dark matter and the number of fermion families all have a common origin.
The breaking of the Peccei-Quinn symmetry sets the scale $\vev{\sigma}$ for the type-I Dirac seesaw mechanism responsible for neutrino masses, see Fig.~\ref{fig} and Eq.~\ref{ssm2}. 
We compare our axion properties with those of simpler axion setups that do not address neutrino mass generation. 
 Our revamped axion mimics the conventional DFSZ and KSVZ axions, as it couples both to SM quarks, and to the new exotic quarks required to realize the 
 3-3-1 symmetry structure.
 However, as seen in Fig.~\ref{plot:agg}, in our scheme the axion has an enhanced coupling to photons.
 In Figs.~\ref{fig:aee}, Fig.~\ref{fig:an} and~Fig.~\ref{fig:ap}
we compare the axion-electron,
axion-neutron and axion-proton couplings with those of the traditional models.
Besides all these differences, our revamped axion behaves as a flavored axion, having flavor-changing couplings to fermions, despite the fact that we do not impose any family symmetry. 
Indeed, this feature leads to 
interesting phenomenological consequences, including flavor-changing axion-emitting two-body meson decays, as seen in 
Fig~\ref{plot:brs}.
We found correlations between the B and D decay branching fractions with that of the most tightly restricted mode, i.e. $\mathcal B (K^+ \to \pi^+ a)$. The latter is displayed in Fig.~\ref{plot:KpiaVsfa} in terms of the axion mass.

Upcoming projected experiments should be able to probe our revamped
axion scenario in a meaningful manner.
Moreover, our scheme also implies a plethora of new signatures associated to the 3-3-1 gauge symmetry structure, required in order to justify the validity of the Dirac neutrino seesaw mechanism. These would bring in new physics phenomena at accessible scales. 

\acknowledgements 
\noindent

AK thanks Avelino Vicente for technical discussions regarding the use of different packages. 
This work is funded by Spanish grants PID2023-147306NB-I00, PID2020-114473GB-I00 and PID2023-146220NB-I00 and by Severo Ochoa Excellence grant CEX2023-001292-S (AEI/10.13039/501100011033) and by Prometeo CIPROM/2021/054 and Prometeo/2021/071 (Generalitat Valenciana).
\\

\bibliographystyle{utphys}
\bibliography{bibliography}
\end{document}